\documentclass[14pt]{article}
\pdfoutput=1
\usepackage{amsmath,amssymb,amsfonts,amsthm,bm,bbm,cancel,wasysym}
\usepackage{epsfig,graphics,graphicx,epstopdf}
\usepackage{array,booktabs,colortbl,colordvi,multirow}
\usepackage{colordvi,color,xcolor}
\usepackage{hyperref}
\usepackage{rotating}

\usepackage{verbatim}
\usepackage{cite}
\usepackage{subfig}
\usepackage{setspace}
\usepackage{url}
\usepackage[percent]{overpic}
\usepackage{slashed}
\usepackage{authblk}
\usepackage{xspace}
\usepackage{fullpage}
\usepackage{hyperref}

\def\ie{{\it i.e.}}
\def\eg{{\it e.g.}}
\def\etc{{\it etc}}

\def\to{\rightarrow}

\newskip\zatskip \zatskip=0pt plus0pt minus0pt
\def\matth{\mathsurround=0pt}
\def\lsim{\mathrel{\mathpalette\atversim<}}
\def\gsim{\mathrel{\mathpalette\atversim>}}
\def\atversim#1#2{\lower0.7ex\vbox{\baselineskip\zatskip\lineskip\zatskip
  \lineskiplimit 0pt\ialign{$\matth#1\hfil##\hfil$\crcr#2\crcr\sim\crcr}}}

\begin{document}


\begin{flushright}
SLAC-PUB-17714\\
\today
\end{flushright}
\vspace*{5mm}

\renewcommand{\thefootnote}{\fnsymbol{footnote}}
\setcounter{footnote}{1}

\begin{center}

{\Large {\bf Towards UV-Models of Kinetic Mixing and Portal Matter III: Relating Portal Matter and R-H Neutrino Masses}}\\

\vspace*{0.75cm}

{\bf Thomas G. Rizzo}~\footnote{rizzo@slac.stanford.edu}

\vspace{0.5cm}

{SLAC National Accelerator Laboratory}\ 
{2575 Sand Hill Rd., Menlo Park, CA, 94025 USA}

\end{center}
\vspace{.5cm}


\begin{abstract}
\noindent  

The kinetic mixing (KM) of a dark photon (DP) with the familiar one of the Standard Model (SM) requires the existence of a new set of fields, called portal matter (PM), which carry both SM 
and dark sector quantum numbers, some whose masses may lie at the TeV scale. In the vanilla KM model, the dark gauge group is just the simple $G_{Dark}=U(1)_D$ needed to describe the 
DP while the SM gauge interactions are described by the usual $G_{SM}=SU(3)_c\times SU(2)_L\times U(1)_Y$. However, we need  to go beyond this simple model to gain a better 
understanding of the interplay between $G_{SM}$ and $G_{Dark}$ and, in particular, determine how they both might fit into a more unified construction. Following our previous analyses, this 
generally requires $G_{Dark}$ to be extended to a non-abelian group, \eg, $SU(2)_I\times U(1)_{Y_I}$, under which both the PM and SM fields may transform non-trivially. In this paper, 
also inspired by our earlier work on top-down models, we consider extending the SM gauge group to that of the Left-Right Symmetric Model (LRM) and, in doing so, through common vacuum 
expectation values, link the mass scales associated with the breaking of $G_{Dark}\to U(1)_D$ and the PM fields to that of the RH-neutrino as well as the heavy gauge bosons of the LRM. 
This leads to an interesting interplay between the now coupled phenomenologies of both visible and dark sectors at least some of which may be probed at, \eg, the LHC and/or at the future 
FCC-hh.
\end{abstract}

\vspace{0.5cm}
\renewcommand{\thefootnote}{\arabic{footnote}}
\setcounter{footnote}{0}
\thispagestyle{empty}
\vfill
\newpage
\setcounter{page}{1}



\section{Introduction and Background}

Although strong evidence for Dark Matter (DM) is known to exist over many length scales, its fundamental nature remains a great mystery. In particular, the answer to the question as to 
just how or if DM may interact with the fields of the Standard Model (SM), apart from via the obvious gravitational interactions, is most pivotal in our attempt to understand how the DM 
may have achieved the relic density determined by Planck\cite{Planck:2018vyg}. More than likely, some new non-SM force(s) must exist to help achieve this result and one might ask  
how such new forces and the familiar ones of the SM may be related (if at all) and if some unified interaction framework might be contemplated. 

Of course these are not new questions and the 
searches for the `traditional' DM candidates, such as the QCD axion\cite{Kawasaki:2013ae,Graham:2015ouw,Irastorza:2018dyq} and weakly interacting massive particles, \ie, 
WIMPS\cite{Arcadi:2017kky,Roszkowski:2017nbc}, continue to push deeper and wider into parameter space with ever greater sensitivities. So far, however, these searches by direct or 
indirect detection 
experiments, as well as those at the LHC\cite{LHC,Aprile:2018dbl,Fermi-LAT:2016uux,Amole:2019fdf,LZ:2022ufs}, have produced negative results, thus excluding an ever growing region of the 
corresponding allowed model space. Over the last few years, the long wait for convincing axion and/or WIMP signatures has led to an ever expanding set of new ideas for the 
nature of DM and its 
interactions with the SM. In particular, it is now clear that both DM masses and coupling strengths to (at least some of) the fields of the SM can both span extremely large  
ranges\cite{Alexander:2016aln,Battaglieri:2017aum,Bertone:2018krk,Cooley:2022ufh,Boveia:2022syt,Schuster:2021mlr} thus requiring a wide variety of very broad and very deep searches. 
In addition, the types of 
interactions that are possible between the SM and DM fields have also been found to be quite numerous and a very useful classification tool to describe these potential structures is via 
renormalizable (\ie, dimension $\leq 4$) and  non-renormalizable (\ie, dimension $> 4$) `portals'. This approach posits the existence of a new set of mediator fields which link the SM to the  
DM and also possibly to an enlarged, potentially complex, dark sector of which the DM itself is its lightest, stable member due to the existence of some new at least approximately 
conserved quantum number.

Of the various portals, one that has gotten significant attention in the literature due to its parameter flexibility is the renormalizable kinetic mixing (KM)/vector 
portal\cite{KM,vectorportal,Gherghetta:2019coi} scenario 
which is based upon the existence of a new dark gauge interaction. One finds that in such a setup, even in its simplest manifestation and for a suitable range of parameters, that 
this scenario allows DM to reach its measured abundance via the usual WIMP-like thermal freeze-out mechanism\cite{Steigman:2015hda,Saikawa:2020swg}. However, this now occurs 
for sub-GeV DM masses by employing this 
new non-SM dark gauge interaction that so far have evaded detection. This simplest and most familiar of these manifestation assumes only the existence of a new $U(1)_D$ gauge group, 
with a gauge coupling $g_D$, under which the SM fields are neutral, thus carrying no dark charges, \ie, having $Q_D=0$, and where the new $U(1)_D$ gauge boson is referred to as the 
`dark photon' (DP) \cite{Fabbrichesi:2020wbt,Graham:2021ggy}, which we will henceforth denote by $V$ or $A_I$ depending on context. 
As noted, to obtain the observed relic density by thermal means, this new $U(1)_D$ is usually assumed to be spontaneously broken at or below 
the $\sim$ few GeV scale so that both the DM and DP will have comparable masses. This symmetry breaking is usually accomplished via the (sub-)GeV scale vev(s) of at least one 
new scalar, the dark Higgs, in complete analogy with the symmetry breaking occurring in the SM. Within such a framework, the the interaction between the SM and the dark sector 
is generated via renormalizable kinetic mixing (KM) at the 1-loop level between the $U(1)_D$ and the SM $U(1)_Y$ gauge fields. The strength of this KM interaction is then described by a 
small, dimensionless 
parameter, $\epsilon$. For this mixing to occur, these loops must arise from a set of new matter fields, usually being vector-like fermions (and/or complex scalars), here called Portal Matter 
(PM) \cite{Rizzo:2018vlb,Rueter:2019wdf,Kim:2019oyh,Rueter:2020qhf,Wojcik:2020wgm,Rizzo:2021lob,Rizzo:2022qan,Wojcik:2022rtk,Rizzo:2022jti,Rizzo:2022lpm,Wojcik:2022woa,
Carvunis:2022yur,Verma:2022nyd}, that, unlike the SM fields, carry {\it both} SM hypercharge (plus other model-dependent SM quantum numbers) as well as a $U(1)_D$ dark charge. 
Subsequent to field redefinitions that bring us back to canonically normalized fields, after both the SM and $U(1)_D$ gauge symmetries are broken, and further noting the large ratio of the 
resulting $Z$ to $V$ masses, this KM leads to a coupling of the DP to SM fields of the form $\simeq e\epsilon Q_{em}$, up to correction terms or order $m_V^2/m_Z^2<<1$. The size of the 
parameter $\epsilon$ is constrained by phenomenology to very roughly lie in the $\epsilon \sim 10^{-(3-4)}$ range given the DM/DP lies within the sub-GeV mass region that leads to the thermal 
DM freeze-out mechanism of interest to us here.  One also finds that in such a setup, for $p-$wave annihilating DM or for pseudo-Dirac DM with a sufficient mass splitting, the rather tight 
constraints arising from the CMB can also be rather easily avoided\cite{Planck:2018vyg,Slatyer:2015jla,Liu:2016cnk,Leane:2018kjk} for a similar range of parameters.

In the conventionally chosen normalization\cite{KM,vectorportal}, with $c_w=\cos \theta_w$, $\epsilon$ can be expressed in terms of the properties of the PM fields themselves appearing 
in these vacuum polarization graphs and is given by the sum
\begin{equation}
\epsilon =c_w \frac{g_D  g_Y}{24\pi^2} \sum_i ~\eta_i \frac{Y_i}{2}  N_{c_i}Q_{D_i}~ ln \frac{m^2_i}{\mu^2}\,,
\end{equation}
where $g_{Y,D}$ are the $U(1)_{Y,D}$ gauge couplings and $m_i(Y_i,Q_{D_i}, N_{c_i})$ are the mass (hypercharge, dark charge, number of colors) of the $i^{th}$ PM field. Here, 
we note that $\eta_i=1(1/2)$ if the PM particle is a chiral fermion (complex scalar) and the SM hypercharge is here normalized so that the electric charge is given as $Q_{em}=T_{3L}+Y/2$. 
It is important to note that in a somewhat more complex scenario where this effective theory is embedded into a broader UV-complete setup, such as we will describe below, this same 
group theory requires that the sum (for fermions and scalars separately)
\begin{equation}
\sum_i ~\eta_i \frac{Y_i}{2} N_{c_i} Q_{D_i}=0\,,
\end{equation}
so that one finds that $\epsilon$ is both finite and, if the PM masses and couplings were also known, completely determined within the more fundamental underlying model.  

Clearly it is advantageous to go beyond this rudimentary effective theory to further our understanding of how this (apparently) simple KM mechanism fits together in a single picture with the 
SM, something that we have begun to examine in pathfinder mode employing various bottom-up and top-down approaches in a recent series of papers 
\cite{Rizzo:2018vlb,Rueter:2019wdf,Rueter:2020qhf,Wojcik:2020wgm,Rizzo:2021lob,Rizzo:2022qan,Wojcik:2022rtk,Rizzo:2022jti,Rizzo:2022lpm,Wojcik:2022woa}. Two specific features 
of our general framework are the extension of the $U(1)_D$ dark abelian symmetry to, \eg, the non-abelian, SM-like $G_{Dark}=SU(2)_I\times U(1)_{Y_I}$\cite{Rueter:2019wdf} 
gauge symmetry\cite{Bauer:2022nwt} and the appearance of at least some of the SM fields in common $SU(2)_I$ representations with the PM fields. In such setups, the PM masses 
are themselves generally 
the result of the  $G_{Dark} \to U(1)_D$ symmetry breaking and so, with $O(1)$ Yukawa couplings, will share a similar overall scale with the new, heavy, $Q_{em}=0$, gauge 
bosons, denoted by $W_I^{(\dagger)},Z_I$, associated with the broken group generators. This was seen quite explicitly in Ref.\cite{Rueter:2019wdf} whose PM content and the 
$G_{Dark}=SU(2)_I\times U(1)_{Y_I}$ gauge group were both inspired by $E_6$\cite{e6,Hewett:1988xc}. One can also consider classes of models wherein the SM gauge group, $G_{SM}$, is 
itself extended as was suggested by the top-down analysis in Ref.\cite{Rizzo:2022lpm}. In any such setup where the SM lepton doublets and PM fields are found to lie in common 
representations of that group, it is easily imagined that there may exist 
a possible relationship between the see-saw mass scale that is responsible for generating small Majorana neutrino masses and that associated with the breaking of $G_{Dark}$ down to 
$U(1)_D$ and producing the PM masses. This is perhaps most easily realized in scenarios loosely described by the product of gauge groups  $G=G_{SM}\times G_{Dark}$ under which 
the DM is a $G_{SM}$ singlet and $G_{SM}$ already naturally gives rise to a see-saw mass structure. The most simple, obvious and familiar example of such a possibility is to 
consider identifying the augmented 
$G_{SM}$ with the Left-Right Symmetric Model (LRM)\cite{Pati:1974yy,Mohapatra:1974hk,Mohapatra:1974gc,Senjanovic:1975rk,book} wherein the usual SM is extended to 
$G=SU(3)_c\times SU(2)_L\times SU(2)_R \times U(1)_{B-L}$, which also has other advantages, \eg, in making it easier to satisfy anomaly constraints and in obtaining a finite 
and calculable value of $\epsilon$ as described above. This is the scenario that we will consider below. As in earlier work, we will employ 
$G_{Dark}=SU(2)_I\times U(1)_{Y_I}$\cite{Rueter:2019wdf} as the simplest non-abelian example 
which can contain an unbroken $U(1)_D$ and which allows for $Q_D=0$ SM fields to lie in common representations with PM fields which must have $Q_D\neq 0$. Of course, this choice 
is hardly unique but its simplicity allows us to more clearly see the relationship between  the right-handed neutrino/see-saw mass scale and that of PM, where $G_{Dark}$ also breaks. 

In order to satisfy our various model building requirements, it is far simpler (and more easily UV-completed) to base our model on a simplified, single-generation version of the one 
appearing in Ref.\cite{Wojcik:2020wgm} (since here we will not be addressing any flavor issues) in the following discussion. Specifically, we'll be considering an extended version of the 
Pati-Salam(PS)-Left-Right Model (LRM)\cite{Pati:1974yy} augmented by a non-abelian dark sector gauge group as discussed in Ref.\cite{Wojcik:2020wgm}, \ie, 
$G=SU(4)_c\times SU(2)_L\times SU(2)_R \times SU(2)_I\times U(1)_{Y_I}$, which we will denote for brevity as $4_c2_L2_R2_I1_{Y_I}${\footnote {As we will see below, we will on some  
occasions refer to the product $2_L2_R$ as simply $2_12_2$ whenever we need to avoid confusion with respect to the fermion assignments.}}  Here it will be assumed that the breaking of 
$SU(4)_c \to SU(3)_c\times U(1)_{B-L}$ occurs at a very large mass scale, $M_c \gsim 10^6$ TeV or even at the unification scale, so that although it determines the initial 
representation structure of the fermion sector 
necessary for anomaly cancellation, \etc, it will not have any phenomenological impact on the discussion that follows below\cite{Dutka:2022lug}. This setup then implies that at accessible 
energy scales below $\sim 10$'s of TeV, the effective gauge group for our discussion is actually just 
$G_{eff}=3_c1_{B-L} 2_L2_R2_I1_{Y_I}$. In such a framework, it is now the two $U(1)$'s, $1_{B-L}1_{Y_I}$, which will undergo KM. While this KM will still manifest itself as a DP which 
(weakly) couples like the SM photon at the $\lsim 1$ GeV mass scale, additional coupling terms to SM fields of various kinds will also be present due to, \eg, mass mixing from the several steps 
of symmetry breaking which are necessary before the $\lsim 1 $ GeV scale is reached and due to the fact that some of the Higgs fields with vevs will carry both SM and non-zero values of 
$Q_D$. As we will see below, 
to maintain anomaly freedom in such a setup, {\it every} SM field is now accompanied by two sets of PM fields with similar SM electroweak quantum numbers, which will together form a complete 
SM/LRM vector-like family, but whose members will transform differently under $G_{Dark}$. As we will see in the analysis that follows, the masses of the PM fields, the mass of the right-handed 
neutrino and the breaking scale for all of the new heavy gauge bosons will all become correlated, intertwining the physics of the LRM and dark sector gauge groups and leading to a complex 
phenomenological structure some of whose implications we will begin to study below. For example, we will find that both of the heavy neutral Dirac PM fields in the model will be split 
into pairs of pseudo-Dirac states via Majorana mass terms arising from some of the same Higgs fields that are responsible for LRM-like neutrino mass generation.

The outline of this paper is as follows:  Following the present Introduction and Background discussion, in Section 2 a broad outline of the model framework will be presented to set 
the overall stage for the analysis that follows. Section 3 will then individually examine the various sectors of this setup, \ie,  the generation of the Dirac and Majorana fermion 
masses together with the corresponding mixings between the PM and SM/LRM fermion fields. The KM and gauge symmetry breaking which takes place in several distinct steps at a hierarchy 
of mass scales and resulting gauge boson masses and mixings that will be important at the electroweak scale and below will then be discussed. An examination of a sample of some of 
the phenomenological implications and tests of this scenario will also be presented along the way throughout this Section as part of the model development, although much of this model still remains to be explored in future work. A summary, a discussion of our results, possible future avenues of exploration and our subsequent conclusions can then be found in Section 4.


\section{Model Setup and Framework}

For our study below, the specific model building requirements will be taken to be as follows: ($i$) Due to the dual quark-lepton and left-right symmetries of the Pati-Salam setup, all of the SM 
fermions will have the need of PM partners. ($ii$) The PM fermions, though vector-like with respect to the SM/LRM gauge groups, should obtain their masses at the $G_{Dark}$ and/or the 
$G_{LRM}$ breaking scale. The combination of ($i$) and ($ii$), in fact, implies that there are now {\it two} distinct PM chiral partners, together forming a single vector-like fermion, for 
each SM field as we will 
see below.  ($iii$) With an eye toward a possible unification in an even larger gauge structure, this setup must be automatically anomaly-free and yield a finite and calculable value for $\epsilon$ 
as described above. These conditions follow automatically from the discussion in Ref.\cite{Wojcik:2020wgm} when the additional family symmetry group is suppressed as will be the case below. 
Some additional constraints associated with the symmetry breaking hierarchy will be subsequently encountered as we move forward with our discussion. 

In terms of the $4_c2_L2_R2_I1_{Y_I}$ gauge groups discussed above and denoting the quantum numbers of the fields by $(4_c2_L2_R2_I)_{Y_I/2}$, 
a single fermion generation, here denoted by $\cal F$, will consist of the following set of fields\cite{Wojcik:2020wgm}:
\begin{equation}
{\cal F}=A(4,2,1,2)_{-1/2}+B(4,1,2,2)_{-1/2}+C(4,2,1,1)_{-1}+D(4,1,2,1)_{-1}\,,  
\end{equation}
and, recalling that under the breaking $SU(4)_c \to SU(3)_c\times U(1)_{B-L}$ at the very large mass scale, $M_c$, assumed here, one has $4\to 3_{1/3}+1_{-1}$. Thus we see that while 
the familiar SM fermions and the RH-neutrino, $f_{L,R}$, which form the usual LRM doublets under $SU(2)_{L(R)}$, lie in the representations $A,B$, additional vector-like (with 
respect to the SM/LRM) fermions, here denoted as $F_{L,R},F_{L,R}'$, are also present. While $(f,F)^T_{L,R}$ combine to form the $2_I$ doublets, $A,B$, respectively, $F_R'$ and $F_L'$, 
are both $2_I$ singlets that form the corresponding representations $C$ and $D$. It is important to note that $F_R'$ is a $2_1(=2_L)$ doublet and a $2_2(=2_R)$ singlet while the reverse 
is true for $F_L'$. It is this subtle and somewhat unusual 
fermion assignment which prevents us from completely casually referring to $2_12_2$ as $2_L2_R$ in the usual manner, though we will with some caution mostly employ this later notation to 
make contact with the traditional LRM setup as long as the careful Reader is always mindful of the subtleties involved. Within this framework, we see explicitly that the SM fermion fields will all 
have $Q_D=0$ while the PM fermions will all have $Q_D=-1$. In such a setup with, \eg, $SU(2)_I$ acting vertically and $SU(2)_L$ acting horizontally, the LH SM fermions will appear 
as bi-doublets with their unprimed PM partners, \eg, 
\begin{equation}
\begin{pmatrix} \nu_L & e_L \\ N_L & E_L \\ \end{pmatrix}, ~\begin{pmatrix} u_L & d_L \\ U_L& D_L \\  \end{pmatrix}\,,
\end{equation} 
and similarly for the RH states, while the primed PM states will be appear purely `horizontal' as they are $SU(2)_I$ singlets but $SU(2)_L$ or $SU(2)_R$ doublets, \eg, $(N',E')_{R,L}$, 
respectively (note the flipped helicities),  and 
$(U',D')_{R,L}$. It is important to note that while transforming quite differently under $G_{Dark}$, with the caveats mentioned above, the (chiral) fields $f,F$ and $F'$ will all have somewhat 
similar transformation properties under the $3_c2_L2_R1_{B-L}$ combination of gauge groups since $F,F'$ together will form a SM vector-like copy of the chiral fermion, $f$.

As seen above, at the level of $G_{eff}=3_c1_{B-L} 2_L2_R2_I1_{Y_I}$, it will be the two $U(1)_{B-L}$ and $U(1)_{Y_I}$ abelian gauge fields which will undergo KM due to the now familiar 
PM loops. Denoting the $B-L$ and $Y_I$ kinetically mixed gauge field strengths as $\tilde B_{\mu\nu},\tilde D_{\mu\nu}$, respectively, the KM piece of the Lagrangian for the 
two $U(1)$'s above the $2_R2_I$ breaking scales, $M_{R,I}$, can be written as 
\begin{equation}
{\cal L}_{KM} =-\frac{1}{4}\tilde B_{\mu\nu}^2-\frac{1}{4}\tilde D_{\mu\nu}^2+\frac{\sigma}{2}\tilde B_{\mu\nu}\tilde D^{\mu\nu}\,,  
\end{equation}
where the dimensionless parameter, $\sigma \sim 10^{-(3-4)}$, describes the strength of the KM. Below, we will connect this parameter with the familiar $\epsilon$ one of similar magnitude 
which describes the KM of the DP with the SM photon at low energy scales. Employing similar notation to the above, one finds that $\sigma$ is given by 
\begin{equation}
\sigma =\frac{g_{B-L}  ~g_{Y_I}}{24\pi^2} \sum_i ~\eta_i \frac{Y_{I_i}}{2} \frac{(B-L)_i}{2} N_{c_i}~ ln \frac{m^2_i}{\mu^2}\,,
\end{equation}
with $g_{B-L,Y_I}$ being the $U(1)_{B-L,Y_I}$ gauge couplings with their associated quantum numbers and that, correspondingly, the requirement
\begin{equation}
\sum_i ~\eta_i \frac{Y_{I_i}}{2} \frac{(B-L)_i}{2} N_{c_i} =0\,,
\end{equation}
so that the requirement that $\sigma$ is finite and calculable is indeed found to be satisfied for the fermion content of the setup above. We will later see that this remains true once the 
scalar degrees of freedom are 
included below as these are just `products' of the above fermion representations and so will just have the quantum numbers which are either sums and differences of those of the $A-D$ 
fermion fields which themselves lead to a finite $\sigma$.  As is usual, since the KM parameter (in this case $\sigma$) is expected to be so small, we can safely work to linear order in 
this parameter {\it most} of the time and so we observe that the KM above is removed by the familiar field redefinitions 
$\tilde B_{\mu\nu} \to B_{\mu\nu}+\sigma D_{\mu\nu}$, $\tilde D_{\mu\nu} \to D_{\mu\nu}$ and leads to the following interaction structure (in obvious notation) 
\begin{equation}
g_{B-L} \frac{B-L}{2} B_\mu+\Big(g_{Y_I} \frac{Y_I}{2}+\sigma g_{B-L}\frac{B-L}{2}\Big) D_\mu\,,  
\end{equation}
where $B,D$ here are simply the associated canonically normalized gauge fields, and which then will appear as one of the pieces of the covariant derivative. 

For the neutral, hermitian fields (apart from QCD which remains exactly as in the SM), the part of the covariant derivative describing interactions can be 
suggestively written in the familiar $G_{SM/LRM}  \times G_{Dark}$ (from Ref.\cite{Rueter:2019wdf}), but not-quite mass eigenstate, basis as (suppressing Lorentz indices) 
\begin{equation}
\begin{aligned}
{\cal L}^h_{int}=&~eQA+\frac{g_L}{c_w}\Big(T_{3L}-xQ\Big)Z+\frac{g_L}{c_w}[\kappa^2-(1+\kappa^2)x]^{-1/2}\Big(xT_{3L}+\kappa^2(1-x)T_{3R}-xQ\Big)Z_R\\&+\frac{g_I}{c_I}
\Big(T_{3I}-x_IQ_D\Big)Z_I+g_DQ_DA_I+\sigma \lambda g_L \frac{B-L}{2}\big(c_IA_I-s_IZ_I \big),
\end{aligned}
\end{equation}
with $Q=Q_{em}$ and, more suggestively, with the replacement $V\to A_I$ to further heighten the analogy to the SM.  
Here, we've introduced the usual SM relationship $e=g_Ls_w$ with $s_w(c_w)=\sin \theta_w(\cos \theta_w)$, \etc, as well as the abbreviations $x=x_w=s_w^2$, $\kappa=g_R/g_L$ and also 
\begin{equation}
\lambda^2=\frac{\kappa^2x}{\kappa^2-(1+\kappa^2)x}=\kappa^2 x \Omega^{-2}\,,  
\end{equation}
so that $g_{B-L}=g_L\lambda$; note the constraint arising from the requirement of real couplings in that $\kappa$ is bounded from 
below, \ie, $\kappa^2>x/(1-x)=t_w^2$\cite{Rizzo:2006nw} so that $\kappa \gsim 0.55$. 
Similarly, in close analogy to the SM and as we've employed in earlier work\cite{Rueter:2019wdf}, we've also defined $g_D=g_Is_I=e_I$, with $s_I$ being the 
analog of $s_w$, \etc,  to be the $U(1)_D$ gauge coupling of the light DP, together with $x_I=s_I^2$ and $Q_D=T_{3I}+Y_I/2$ as the usual dark charge to which the DP couples, again 
simply completely paralleling the SM case. We note that $Z_I$ in this setup couples universally, \ie, independent of flavor or generation, for all choices of $f$, $F$ and $F'$, before the effects 
of fermion mixing are included as we will discuss below. The last term in the above expression is the one arising from the KM $\sigma g_{B-L}\frac{B-L}{2}D$ coupling term in Eq.(8) 
above but is now written here more suggestively in terms of the  $Z_I,A_I$ fields. Similarly, the interactions of the {\it non-hermitian} gauge bosons, $W_{L,R,I}$, are controlled by the gauge 
group coupling structure which in this same approximate mass eigenstate basis is given by 
\begin{equation}
{\cal L}^{nh}_{int}=~\frac{g_L}{\sqrt 2}\big(T_L^+W+{\rm{h.c.}}\big) +(L\to R)+\frac{g_I}{\sqrt 2}\big(T_I^+W_I+\rm{h.c.}\big),
\end{equation}
where $T_{L,R,I}^{+(-)}$ are the corresponding isospin raising(lowering) operators for the $SU(2)_{L,R,I}$ gauge groups, respectively.

\section{Analysis and Phenomenology}

To go further, we must address how the various symmetries are broken and how the corresponding gauge bosons and fermions obtain their different masses. In this construction, ignoring $M_c$, 
there are (at least) 3 distinct, widely separated mass scales. At the highest mass scales, $\gsim 10$'s of TeV, the LRM must break down to the SM (at $M_R$) and also 
$G_{Dark} \to U(1)_D$ (at $M_I$), both of which may be related as we will discuss below. At the $\sim 100$ GeV scale, the SM undergoes the familiar electroweak symmetry breaking 
while $U(1)_D$ itself breaks at low energies, \ie, $\sim 1$ GeV or below. Note the hierarchy of roughly a factor of $\sim 10^2$ between these three scales so that it is reasonable to 
consider them somewhat sequentially and we note that $Q_D$ will be a conserved quantity until quite low scales are reached. Thus the vevs of the 
Higgs fields which are {\it mainly} responsible for the first two symmetry breaking steps can only arise from neutral scalar multiplet members also having $Q_D=0$. This will be important 
to remember in the following discussion.

\subsection{Dirac Fermion Masses}

The quantum numbers of the active set of Higgs fields, $H_{1-4}$, all assumed to be color singlets and having $B-L=0$, that are needed to generate the various Dirac fermion masses are 
easily obtained by taking appropriate products of the fermion representations $A-D$ above and can be expressed, ie, via the Yukawa couplings which generalizes the usual LRM structure as:
\begin{equation}
{\cal L}_{Dirac} =\bar A_L B_R(y_1H_1+\tilde y_1 \tilde H_1)+y_2\bar A_L C_RH_2+y_3\bar D_L B_RH_3+\bar D_L C_R (y_4 H_4+\tilde y_4 \tilde H_4)+\rm{h.c.}\,
\end{equation}
where as usual $\tilde H_i =i\sigma_2 H_i^*\sigma_2$, with $\sigma_2$ being the Pauli matrix, and where the $y_i, \tilde y_i$ are Yukawa couplings, so that the 
$H_i$'s $(2_L,2_R,2_I)_{Y_I/2}$ quantum numbers can be easily chosen to be  
\begin{equation}
H_1(2,2,1)_0,~ H_2(1,1,2)_{1/2},~H_3(1,1,2)_{-1/2},~H_4(2,2,1)_0\,. 
\end{equation}
Note that we will not necessarily impose any $P,C$ or $CP$ symmetries as might be the case in the usual LRM on these Yukawa couplings in the discussion below but for simplicity 
alone we will assume that all 
of the couplings and vevs are real.  For the immediate discussion, we will focus ourselves only on the Higgs vevs which do {\it not} break $U(1)_D$ and so all correspond to $Q_D=0$ 
components of the $H_i$. The effects of any small additional terms due to possible 
$Q_D\neq 0$ vevs can be added later on as a perturbation upon those which we will now discuss as these are relatively quite highly suppressed by factors of (at least) $10^2$.   
Note that we will treat $H_1$ and $H_4$, which are typical LRM bi-doublets, as distinct fields and we will not take $H_2$ and $H_3$ to be the complex conjugates of each other so that they too 
are also unrelated fields.  Note further that the two vevs contained in each of $H_{1,4}$ are of the electroweak scale while the single vev in each of $H_{2,3}$ will be at the $\sim 10$ TeV scale 
or so and these will lead to the breaking of $SU(2)_I\times U(1)_{Y_I} \to U(1)_D$ as will be discussed later below. 

Denoting the generic set of weak eigenstate fermion fields as ${\cal F}^0_{L,R}=(f,F,F')^0_{L,R}$ in the notation employed above, the vevs within the $H_i$ will then generate a 
$3\times 3$ mass matrix of the form 
\begin{equation}
\bar{\cal F}_L^0 ~{\cal M}~{\cal F}_R^0\,, 
\end{equation}
whose entries will depend upon the location of the $Q_D=0$ elements within the various Higgs representations and which can be diagonalized as is usual by a bi-unitary transformation
\begin{equation}
{\cal M}_D =U_L~{\cal M}~U_R^\dagger\,. 
\end{equation}
As noted, ignoring the possibility of $CP$-violation, \etc, we can for simplicity take the elements of ${\cal M}$ to be real so that this $3\times 3$ matrix can be {\it symbolically} (as the 
$2_L2_R$ subspace itself does not appear here) written, after absorbing the various Yukawa couplings into the vevs for brevity, as
\begin{equation}
{\cal M} \sim \frac{1}{\sqrt 2}~\begin{pmatrix}  v & 0 & 0 \\ 0 & v & \Lambda \\0 & \Lambda' & v' \\ \end{pmatrix}\,,
\end{equation} 
where $v,v'$ represent generic weak scale vevs $\sim 100$ GeV, arising from $H_1$ and $H_4$, respectively, and $\Lambda,\Lambda'$ represent vevs at the $\sim 10$ TeV scale, arising 
from $H_2$ and $H_3$, respectively. To clarify, it should be recalled that since both $H_{1,4}$ are standard bi-doublets in the $2_L2_R$ subspace, $v$ and $v'$ here are both just 
symbolic `projections' of the familiar electroweak scale vevs, $k_{1,2}$ and $k'_{1,2}$, that we would usually encounter in the ordinary LRM, where we would instead write the vevs of 
$H_{1,4}$, now in the $2_L2_R$-subspace language as{\footnote {Note that the overall factor of $1/\sqrt 2$ here is associated with each of the vevs appearing in this mass matrix.}}  
\begin{equation}
<H_1>=\frac{1}{\sqrt 2} \begin{pmatrix}  k_1 & 0 \\ 0 & k_2 \\ \end{pmatrix},~<H_4>=\frac{1}{\sqrt 2} \begin{pmatrix}  k'_1 & 0 \\ 0 & k'_2 \\ \end{pmatrix}\,,
\end{equation} 
which, given the coupling structure above, would allow, \eg, different masses for up- and down-type quarks. Thus $v,v'$ can just be thought of as symbolic appropriate linear combinations 
of the $k_i$ and $k'_i$, respectively, depending upon the $2_L2_R$ transformation properties of the relevant fermion field. Next, we can easily determine the bi-unitary transformations 
needed to diagonalize this matrix, which here are just (almost) essentially rotations, $U_{L,R}$, via the standard relation 
\begin{equation}
{\cal M}_D^2=U_L^\dagger {\cal MM}^\dagger U_L= U_R^\dagger {\cal M}^\dagger {\cal M} U_R\,,
\end{equation} 
where ${\cal M}_D$ is the resulting diagonal mass matrix. From the form of ${\cal M}$, and the corresponding products with its hermitian conjugate, it can be seen that the generated mixing 
at this level of symmetry breaking lies totally within the $F-F'$ sector in a $2\times 2$ sub-matrix and that the mass of the unmixed SM field, $f$, is just $\sim y_1(y_1')v/\sqrt 2$, as is similar 
to the LRM/SM{\footnote {Of course in actuality this is really just a weighted sum of the $k_i$ or $k_i'$ vevs.}} . With this and the assumptions we made above we can then replace $U_L$ by a 
simple $2\times 2$ rotation matrix, $O_L$, and $U_R$ by a simple $2\times 2$ rotation together with a discrete transformation, \ie,  $U_R=O_RP$ where $P$ is just 
\begin{equation}
P=\begin{pmatrix} 0 & 1 \\ 1 & 0 \\ \end{pmatrix}\,,
\end{equation} 
and $O_{L,R}$ are each described by a single mixing angle, $\theta_{L,R}$, which are given in terms of ratios of the vevs $v,v', \Lambda, \Lambda'$ by 
\begin{equation}
\tan 2\theta_L=\frac{2(\Lambda v'+\Lambda' v)}{\Lambda^2+v^2-\Lambda'^2-v'^2},~\tan 2\theta_R=\frac{2(\Lambda v+\Lambda' v')}{\Lambda^2+v'^2-\Lambda'^2-v^2}\,,
\end{equation} 
both of which are of similar magnitude, $O(10^{-2})$, given the anticipated mass scales of the various vevs. These mixing angles can then be used to describe how the resulting mass 
eigenstate fermion fields, here termed $f,F_{1,2}$ at this stage of symmetry breaking, will interact with the many gauge bosons in the current setup, in particular, the heavy gauge 
fields associated with the broken $G_{Dark}=SU(2)_I \times U(1)_{Y_I}$ generators. For completeness, we note that to leading order in the squared vev ratios 
$(v^2,v'^2)/(\Lambda^2,\Lambda'^2)$, the mass squared eigenvalues for $F_{1,2}$ are given by the expressions 
\begin{equation}
m_{1,2}^2 \simeq \frac{1}{2}(\Lambda^2+v^2+\Lambda'^2+v'^2) \pm \frac{1}{2}|(\Lambda^2+v^2-\Lambda'^2-v'^2)| \pm \frac{(\Lambda v'+\Lambda' v)^2}{|\Lambda^2-\Lambda'^2|}\,,
\end{equation} 
along the lines that we might have expected, \ie, that essentially $m_1\simeq \Lambda$ while $m_2\simeq \Lambda'$ up to few percent corrections. 
Note that before explicitly evaluating these expressions for the mixing angles and masses, however, we must appropriately restore all the suppressed Yukawa couplings, \eg,  $v\to y_1 (y_1')v$, 
$\Lambda \to y_2\Lambda$, $\Lambda' \to y_3 \Lambda'$, and $v' \to y_4(y_4') v'$.

One very simple but important application of this mixing analysis is to, \eg, identify the PM mass eigenstates sharing the $SU(2)_I$ left- and right-handed doublets together with the SM 
fields, $f_{L.R}$, as these allow the PM states to decay via, \eg, $F_{1,2}\to f W_I$. This is easily done and one finds, defining $c(s)_{L,R}=\cos (\sin) \theta_{L,R}$, that 
\begin{equation}
F_L=F_{1L}c_L-F_{2L}s_L, ~F'_L=F_{2L}c_L+ F_{1L}s_L,~+({\rm L}\to {\rm R}, F\Leftrightarrow F')\,.
\end{equation} 
Since the resulting $fF_{1,2}W_I$ couplings are non-chiral, one possible implication of this is that one-loop graphs can produce a significant effective dipole moment type interaction of the 
SM fermions with $A_I$ at 1-loop that can have important implications for DM searches and associated phenomenology as was discussed via a toy example in Ref.\cite{Rizzo:2021lob} but 
here can realized in a more realistic fashion. Specifically, comparing with this earlier work, one finds the scale associated with these dipole couplings to be given by 
\begin{equation}
\frac{1}{\Lambda_f }=\frac{\alpha_D}{32\pi s_I^2}~\Sigma_i ~\frac{G(y_i)}{m_i}(v_i^2-a_i^2)\,, 
\end{equation}
where we have defined the mass squared ratio $y_i=m_i^2/m_{W_I}^2\sim$ O(1), the loop function $G(y)$ (which numerically is generally also O(1) ) is given by 
\begin{equation}
G(y)=3y^2~\Big[\frac{-2(y-1)+(y+1)~ln(y)}{(y-1)^3}\Big]\,, 
\end{equation}
and the $v_i^2-a_i^2$ factors can be directly obtained from the equations above, \ie, $v_1^2-a_1^2=4c_Ls_R$ and $v_2^2-a_2^2 =-4c_Rs_L$, respectively, so that the $F_{1,2}$ loop 
contributions relatively destructively interfere. For typical choices of the TeV scale PM and $W_I$ masses and associated model parameters, one might then expect to obtain values for 
$\Lambda_f \sim100$'s of TeV in the present setup, which is a phenomenologically interesting range{\footnote {Note that in addition to the $W_I$ contribution to $\Lambda_f$ discussed 
here, there are also potential contributions arising from both CP-even and CP-odd Higgs scalar exchanges which can also yield results of a similar magnitude.}}.

Also, it is interesting to note that since the 2 sets of fermions with $Q_D=-1$, $T_{3I}=-1/2,0$ now mix, the $F_{1,2}$ states will also have 
off-diagonal, `flavor changing neutral current'-like couplings 
to the neutral $Z_I$ gauge boson which will then propagate to the new gauge boson mass eigenstates that we will describe in more detail below. 

Finally, as noted previously, it is important to recall that when/if at least some of the $Q_D\neq 0$ vevs that are possible in the $H_i$ are turned on at much smaller scales below 
$\sim1$ GeV, the mixing as discussed above 
will be slightly perturbed. These new mass terms will be of order $\lsim 10^{-2} (v,v')$ so will not alter the results obtained above 
very significantly in a numerical fashion {\it except} that they will generate $f-F_{1,2}$ mixing, which 
{\it is} phenomenologically important. In particular, we see that both $H_{2,3}$ can have such small, $\lsim 1$ GeV, vevs, $(\lambda,\lambda')/\sqrt 2$, respectively, in obvious notation, that 
will directly couple the LH- and RH-handed components of $f$ and $F'$.  In the $f-F_1-F_2$ basis obtained above, the resulting perturbed, now almost diagonal mass matrix will then appear as 
\begin{equation}
{\cal M}_{D_P} \simeq ~\begin{pmatrix}  m_f & a & b \\ a' & m_1 & 0 \\ b'& 0& m_2 \\ \end{pmatrix}\,,
\end{equation} 
where $(a,b)=y_2 \lambda (c_R,-s_R)/\sqrt 2$ and $(a',b')=y_3 \lambda' (s_L,c_L)/\sqrt 2$. Diagonalization of this matrix produces these $f-F_{1,2}$ mixings and also slight shifts (to 
lowest order in the small parameters) 
the fields such as $f_L\to f_L+aF_{1L}/m_1+bF_{2L}/m_2$, $F_{1L}\to F_{1L}-af/m_1$, \etc, so that $A_I$ can now couple off-diagonally to the $f$ and $F_{1,2}$ mass eigenstates 
in a generally parity violating, yet non-chiral manner, \ie, 
\begin{equation}
-\frac{g_D}{m_1}\bar F_1\gamma_\mu \big(aP_L+a'P_R \big)f A_I^\mu~+(1\to 2, a\to b) +\rm{h.c.}\,,
\end{equation} 
which then allows for the dominant decay paths $F_{1,2}\to fA_I$. Recall that this decay mode is always found to be the most important one for the PM fields in comparison to other decay 
paths generated by such mixings for more conventional vector-like fermions such as $F\to fZ,fH$ or $F\to f'W$. Although {\it all} of these decay modes are apparently 
suppressed by rather small mixing angles and/or mass ratios, the amplitude for the decay into the DP is also {\it enhanced} by large factors of $m_{1,2}/M_{A_I}>>1$ through the 
longitudinal couplings 
of the DP. Numerically, this enhancement can compensate rather completely for the presence of the small mixing angles. In particular, this is quantitatively similar to the 
results found in Refs.~\cite{Rizzo:2018vlb,Rueter:2019wdf} in slightly different contexts and leads to rapid PM decays generated by via the application of the Goldstone Boson 
Equivalence Theorem\cite{GBET} applied in the scalar sector and/or the dominance of the longitudinal modes of the $A_I$ (\ie, the equivalent of the Goldstone boson) since the 
PM fermion masses $m_{1,2}$ are so much larger than that of $A_I$ itself.  In the likely event that the DP appears in collider detectors as MET, the signatures for pair production of these PM 
states will then be observable pairs of SM states, $e^+e^-,\mu^+\mu^-,\bar bb, ~\bar tt$, etc, accompanied by this MET in a manner qualitatively similar to those employed in SUSY 
searches{\footnote {For an overview of the current LHC PM search limits and future prospects, see Ref.\cite{Rizzo:2022qan}; current limits range from 0.9 to 1.5 TeV depending upon the 
PM flavor.}}. 

It should be noted that $W_I,W_I^\dagger$ will also pick up a {\it diagonal} coupling to $\bar f f$ via this same tiny mixing, $\sim (a,a')/m_1 \sim 10^{-4}$, but which in this case is {\it not} offset 
by a large longitudinal enhancement in any decay process and so is not likely to be of much phenomenological relevance in , \eg, the single resonant production of $W_I,W_I^\dagger$ gauge 
bosons at colliders.

\subsection{Neutral Fermion Majorana Masses}

When we consider the $Q_{em}=0$ leptonic components of ${\cal F}$, \ie, $\nu,N,N'$ in the weak basis, they can be `self-coupled' in a manner such that these neutral, neutrino-like 
fields may all obtain Majorana masses from the vevs of suitably chosen Higgs scalars which will carry $|Q_D|=0,1$ or 2, \eg, via the Yukawa structure 
\begin{equation}
{\cal L}_{Majorana} = z^L_1 \bar A^c_L i\sigma_2 A_L \Delta_L +z^L_2 \bar D^c_L i\sigma_2 D_L \tilde \Delta_L +z^L_3 \bar A^c_L i\sigma_2 D_L X_L +(A\to B, D\to C,L\to R)+\rm{h.c.}\,
\end{equation}
where the $z^{L,R}_i$ are new Yukawa couplings, $\sigma_2$ is the Pauli matrix as above and $\Delta_{L,R}, \tilde \Delta_{L,R}$ and $X_{L,R}$ are the 
appropriate Higgs fields, whose active, color-singlet components that will concern us here will now all carry $|L|=2$.~{\footnote{Recall that $F'_{L,R}$ is a doublet under $SU(2)_{R,L}$ and a 
singlet under $SU(2)_{L,R}$.}} The quantum numbers of these Higgs representations in terms of $(2_L,2_R,2_I)_{Y_I/2}$ are easily seen to be just given by 
\begin{equation}
\Delta_L(3,1,3)_1,~ \tilde \Delta_L(3,1,1)_{2},~X_L(2,2,2)_{3/2}, ~+~(L\to R)\,, 
\end{equation}
so that while $\Delta_{L(R)}$ and $\tilde \Delta_{L(R)}$ are $SU(2)_{L(R)}$ isotriplets, $X_{L,R}$ are bi-doublets of $SU(2)_{L,R}${\footnote {Note that, in all generality, we allow $X_L$ and 
$X_R$ to be different scalar fields for this discussion but this need not be the case.}}. Further, $\Delta_{L,R}$  $[\tilde \Delta_{L,R}]$ are 
$SU(2)_I$ triplets [singlets] while $X_{L,R}$ are $SU(2)_I$ doublets. Given these quantum numbers we see that the vevs of all of the neutral component fields contained in any of the 
$\tilde \Delta_{L,R}$ and $X_{L,R}$ will be associated with a {\it non-zero} value of $Q_D$ thus will necessarily lead to a breaking of $U(1)_D$. Thus these vevs must be quite 
small, \ie, certainly $\lsim 1$ GeV or so and will be ignored at the present stage of the discussion but will be returned to below. Meanwhile, only one component in each of 
$\Delta_{L,R}$ has both $Q_{em}=Q_D=0$ and so 
can obtain a vev without breaking $U(1)_D$ and these can be identified with the familiar triplet vevs, $v_{L,R}/\sqrt 2$ ignoring potential phases, commonly appearing in the LRM. As is 
usual in that framework and for all the familiar reasons, \eg, $\rho$ or oblique $T$ parameter constraints\cite{Pati:1974yy,Mohapatra:1974hk,Mohapatra:1974gc,Senjanovic:1975rk,book}, we 
will assume here that $v_L<<v_R$ with $v_R$ setting 
the breaking scale for the LRM, $M_R$. In fact, given such constraints, one might imagine that if $v_L$ is non-zero, its maximum value cannot be too dissimilar from the various possible 
$Q_D\neq 0$ vevs we will later consider below.  As above, we will not explicitly impose any $P,C$ or $CP$ symmetries on these vevs or Yukawa couplings but we will for simplicity of our 
discussion assume that all of them are real. 

One difference between the current setup and the classic LRM scenario, however, is that both of these $Q_D=0$ allowed vevs, $v_{L,R}$, arise from fields which are seen to also be $SU(2)_I$ 
triplets and 
as such these vevs will {\it also} lead to the breaking of $SU(2)_I \times U(1)_{Y_I} \to U(1)_D$. Comparing this to the discussion in the last subsection, we now observe that there are two 
limiting possibilities obtained by comparison of these multi-TeV scale vevs: 
if $v_R>>\Lambda,\Lambda'$ then the $SU(2)_I \times U(1)_{Y_I} \to U(1)_D$ breaking scale is also set by $v_R$ and so $M_I\simeq M_R$. However, if  $v_R<<\Lambda,\Lambda'$, 
then we instead find that  $M_R<<M_I$ and thus it must be so that $M_R\leq M_I$ is always satisfied in the present model setup. Generically, without tuning we might expect to end up 
in the middle of these two extremes so  that all of these vevs are semi-quantitatively comparable and we will treat them in all generality as such in the analysis that follows in the next 
subsection when we discuss the nature of the various gauge bosons masses, \etc, in the current setup.

Given this discussion it is clear that only one large Majorana fermion mass term can be generated at this $Q_D=0$ level and this is due to $v_R$ since all of the other potentially 
contributing vevs are constrained to be very small, at the GeV level or below. This implies that at or above the mass scale of SM electroweak symmetry breaking, $\sim 100$ GeV, the 
neutral fields $N,N'$ will mix as described above to form the two Dirac mass eigenstates, $N_{1,2}$, while $\nu_L-\nu_R$ will form a Majorana mass matrix as in the LRM with the 
conventional see-saw mechanism being active via the hierarchal  $v_L<< v,v' <<v_R $ vevs.  
This decoupled picture will, however, be slightly perturbed once the set of additional, small lepton-number violating,  $Q_D\neq 0$ vevs get turned on. In particular, those associated with the 
$T_{3L(R)}=1$ members of $\tilde \Delta_{L(R)}$, \ie, $\tilde \Delta_{L(R)}^0/\sqrt 2$, and those of the corresponding $T_{3L(R)}=1,T_{3I}=-1$ members of $\Delta_{L(R)}$, \ie, 
$v_{L(R)}''/\sqrt 2$, will turn out to play the dominant roles since these are both $|Q_D|=2$ fields which are obtaining vevs and, as we'll see, produce mass terms that 
also lie along the diagonal of the Majorana mass matrix.

Within this setup, it 
is important to emphasize that while $W_{L,R}$ will couple linear combinations of the $N_i$ with their iso-doublet PM partners, $E_{L,R}$, and, correspondingly $W_I$ will couple them 
to $\nu_{L,R}$, there is no direct $O(1)$ tree-level gauge coupling of the $N_i$ to the usual SM leptons, \eg, $e_{L,R}$. This renders the study of the nature and properties of these interesting 
neutral PM states at colliders somewhat problematic since conventionally we need light charged leptons as decay products and/or co-produced states to probe the \eg, Dirac vs. Majorana 
nature of any new heavy neutral lepton. Heavier gauge bosons that would play this potentially important role would need to live within a larger gauge group, $G'$, within which $SU(2)_I$ and 
$SU(2)_{L/R}$ were unified with the possibilities of such a group to be discussed elsewhere. 

Returning now to the Majorana masses themselves, in the original weak eigenstate basis, \ie, $\nu_L,N_L,N_L'+({\rm L}\to {\rm {R}})$, the full $6\times 6$ Majorana mass matrix for the neutral 
fermions can be symbolically written as 
\begin{equation}
M_{Maj}=\begin{pmatrix} M_L & {\cal M} \\ {\cal M}^\dagger & M_R \\ \end{pmatrix}\,,
\end{equation} 
where ${\cal M}$ is the $3\times 3$ Dirac fermion mass matrix given above in the previous subsection and 
\begin{equation}
M_L = \frac{1}{\sqrt 2} ~ \begin{pmatrix} z^L_1v_L & z^L_1v_L' & z^L_3x_L \\ z^L_1v_L' & z^L_1v_L'' & z^L_3x_L' \\z^L_3x_L & z^L_3x_L' & z^L_2\tilde \Delta_L^0  \\ \end{pmatrix} \,,
\end{equation} 
with the elements of $M_R$ given by the same expressions but with $\rm L \to \rm R$. The not previously mentioned remaining $Q_D \neq 0$ vevs that appear here correspond to the 
$T_{3L(R)}=1,T_{3I}=\pm 1/2$ vevs, $(x_{L(R)},x_{L(R}')/\sqrt 2$, of $X_{L(R)}$ and the $T_{3L(R)}=1, T_{3I}=0$ vevs, $\sim v_{L(R)}'/\sqrt 2$, of $\Delta_{L(R)}$.  Note that in the absence 
of any of the small $|Q_D|=1$ vevs, the SM/LRM fields, $\nu_{L,R}$, and the neutral, $N,N'$, PM sector fields will become completely decoupled. 
After diagonalization, the most important effect of the previously noted $|Q_D|=2$ vevs that live along the diagonal of $M_{Maj}$, to leading 
order in the small vev ratios, is to split both of the heavy $N_{1,2}$ Dirac states into pairs of quasi-Dirac/pseudo-Dirac ones\cite{quasi}, \ie, $N_i \to N_i^\pm$. The corresponding masses,   
$m_i \to m_i \pm \delta m_i/2$, where the $\delta m_i$ are just linear combinations of these four vevs along the diagonal with their associated 
Yukawa couplings.  Note that for $N_i$ masses in the expected range of roughly $m_i \sim 1-10$ TeV, the corresponding {\it fractional} mass splittings will then be expected to be only of order 
$\simeq O(10^{-(3-4)})$ making these splittings quite difficult to discern experimentally since the $N_i$ are not directly connected to the 
charged SM fermions via a direct gauge interaction. By this we mean, as was noted above, that 
while the $W_{L,R}$ gauge boson will connect the $N_i$ to the charged PM fields, $E_{L,R}$, and the $W_I$ gauge bosons will connect them to the $\nu_{L,R}$ fields, there are no 
gauge bosons in this set up that will connect the $N_i$ directly to, \eg,  $e^\pm$ making the Majorana nature of the $N_i$ quite difficult to ascertain. It is important to remember that to leading 
order in the small mixing angles, the $N_i$ will dominantly 
decay as $N_1 \to \nu_L A_I$ and similarly, $N_2 \to \nu_R A_I$, if it is kinematically allowed, which are induced by both fermion and, as we'll see below, gauge boson mixing so that 
the conventional vector-like leptonic decay modes into, \eg, $eW$, are usually quite suppressed. However, if the decay through $\nu_R$ is {\it not} kinematically allowed then the $\nu_L A_I$ 
mode still remains open but this is suppressed by a factor of $\simeq s^2_{L,R} \sim 10^{-4}$ in rate.  Instead, the 3-body decay through a virtual $\nu_R$  and/or $E$ may become 
important and this also leads to a visible final state, \ie, not just missing energy from SM neutrinos and DPs. 

As was discussed in Ref.\cite{quasi}, it was noted that the ratio $\rho=(\delta m_i/\Gamma_i)^2$, where $\Gamma_i$ is the $N_i$ total decay width, is, in principle, an excellent probe of the 
Majorana/Dirac nature of such new neutral heavy leptonic states. From the arguments above, we already expect that $\delta m_i/m_i \lsim 10^{-(3-4)}$ whereas the ratio $\Gamma_i/m_i$ 
arising from the dominant decay into the $\nu_L A_I$ final state is roughly $\lsim 10^{-(2-3)}$ implying that $\rho$ may be relatively small for much of the model parameter space. A more 
detailed analysis of this possibility, however, lies beyond the scope of the current discussion. 

We leave a further study of these effects to future work.

\subsection{Gauge Boson Masses and Mixings}

The couplings of the many gauge bosons to the various fermions introduced above depend not only upon the mixings between these fermions states as already discussed but also on the 
KM and mass mixings among the gauge fields themselves which we will now consider.

The gauge bosons masses and mixings are rather complex in this scenario due to the presence of both KM as well as mass mixing at multiple Higgs vev-induced breaking scales. However, 
since these scales are widely separated by roughly 2 orders of magnitude, we can treat them in stages one at a time in a perturbative manner. It is natural that we will begin this analysis 
by working in the convenient and suggestive basis described by ${\cal L}^h_{int}$ and 
${\cal L}^{nh}_{int}$ above and first consider the effects of the largest vevs, $v_R,\Lambda$ and $\Lambda'$, neglecting the effects of KM, upon the real, hermitian gauge fields. At this level, 
only the $Z_R$ and $Z_I$ fields can obtain masses so that in this $2\times 2$ subspace one obtains a mass squared matrix of the form 
\begin{equation}
M^2_{RI}=(g_Lc_wzv_R)^2~\begin{pmatrix} 1 & \gamma \\ \gamma & \gamma^2(1+R) \\ \end{pmatrix}\,,
\end{equation} 
where
\begin{equation}
z=\frac{\kappa^2}{\sqrt {\kappa^2-(1+\kappa^2)x}}=\kappa^2 \Omega^{-1}, ~~~~\gamma=\frac{g_I/c_I}{g_Lc_wz}, ~~~~R=\frac{\Lambda^2+\Lambda'^2}{4v_R^2}\,, 
\end{equation}
and which can be diagonalized by a rotation to the $Z_{1,2}$ mass eigenstate basis described by an angle 
\begin{equation}
\tan 2\theta_\gamma=\frac{2\gamma}{1-(1+R)\gamma^2}\,,
\end{equation} 
so that $Z_R=c_\gamma Z_1-s_\gamma Z_2$, \etc, in obvious notation. We note that for $R,\gamma \simeq 1$ and $O(1)$ Yukawa couplings, the PM fields, $\nu_R$ and $Z_{1,2}$ all 
will have quite comparable masses in the $\sim$ several TeV range.

\begin{figure}[htbp]
\centerline{\includegraphics[width=4.5in,angle=0]{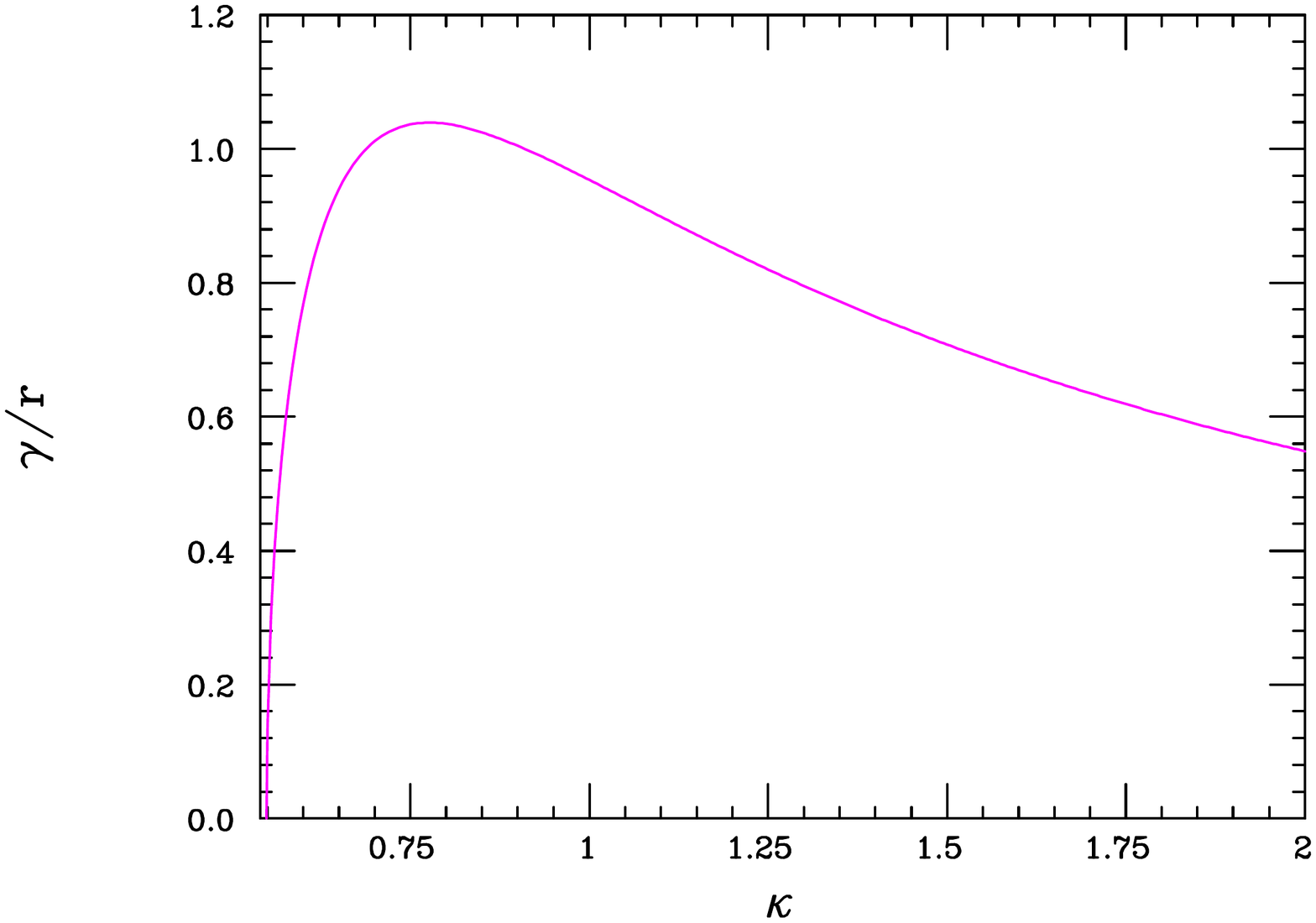}}
\vspace*{-2.0cm}
\centerline{\includegraphics[width=4.5in,angle=0]{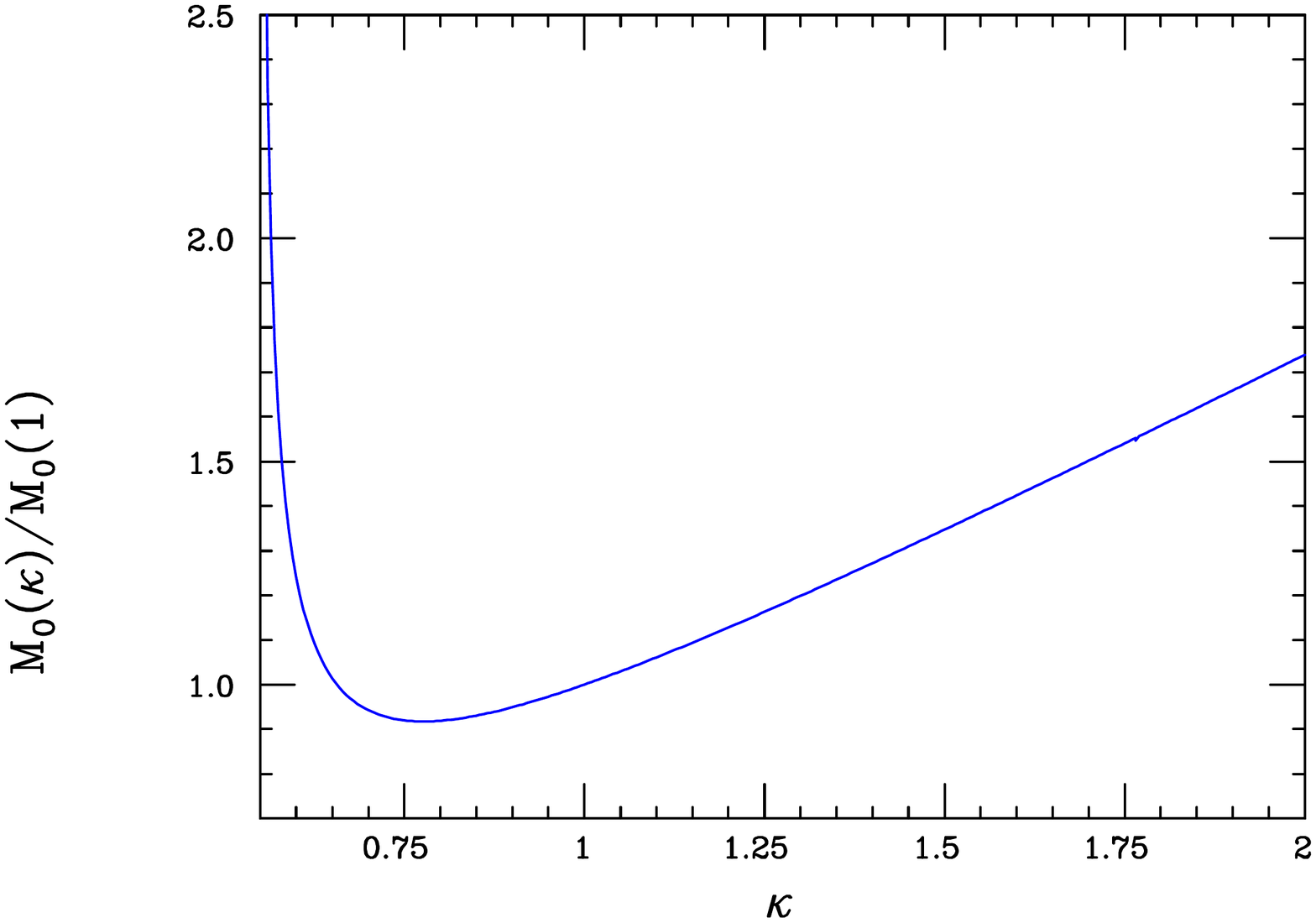}}
\vspace*{-2.0cm}
\centerline{\includegraphics[width=4.5in,angle=0]{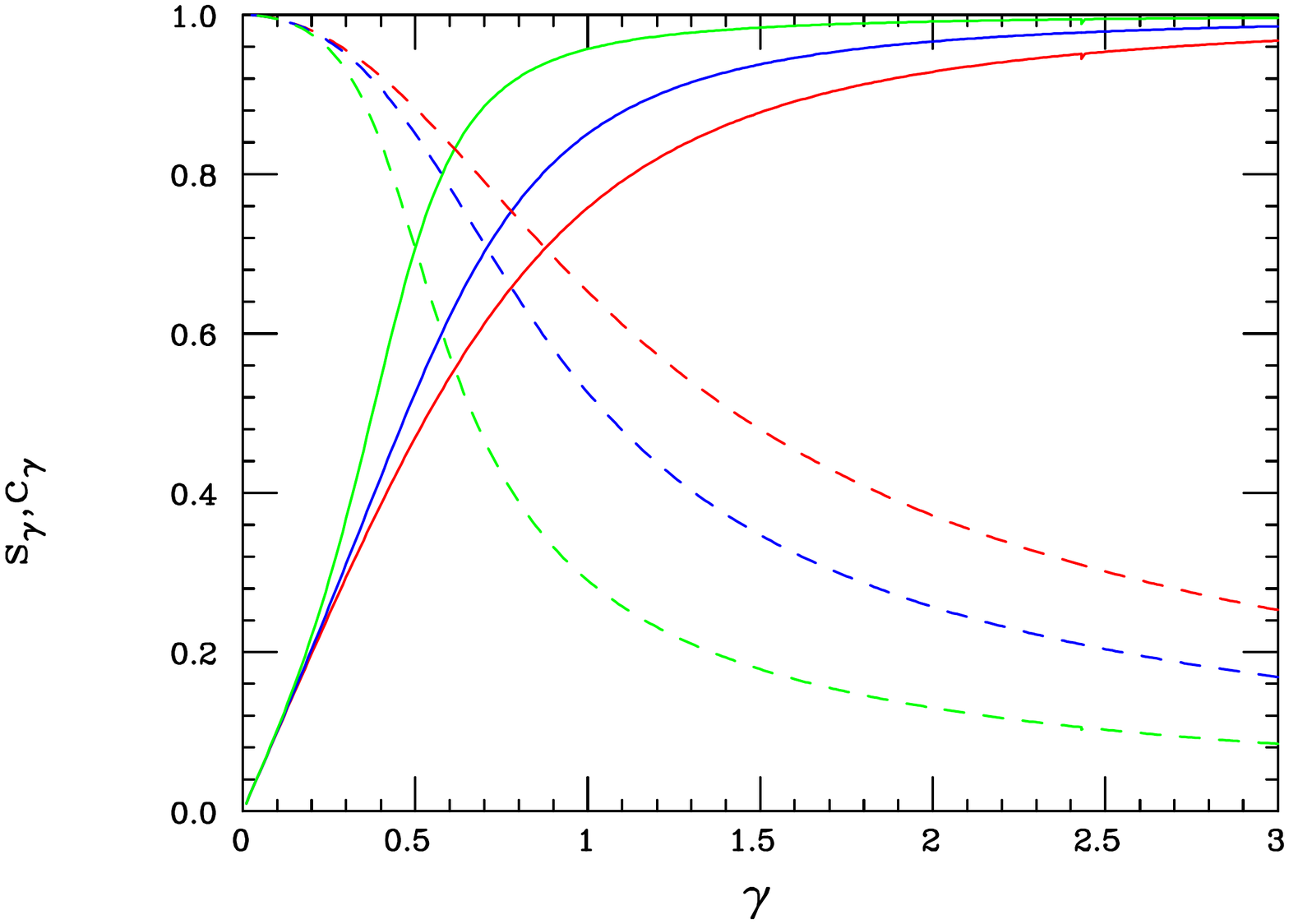}}
\vspace*{-1.3cm}
\caption{The ratios $\gamma/r$ (Top) and $M_0(\kappa)/M_0(1)$ (Middle) as functions of $\kappa$ as described in the text. (Bottom) $s_\gamma$ (solid) and $c_\gamma$ (dashed) 
as functions of $\gamma$ for values of the vev ratio $R$=0.3 (red), 1 (blue) and 3 (green), respectively.}
\label{fig0}
\end{figure}

Note that in terms of, perhaps, the more fundamental quantity, $r=(g_I/c_I)/(g_L/c_w)$, that we employed in our earlier work\cite{Rueter:2019wdf}, and which describes the overall coupling 
strength relative to that of the SM $Z$, the ratio $\gamma/r$ is found to be purely a function of $\kappa$ and is always $\lsim 1.04$ as is shown in the top panel of Fig.~\ref{fig0}. 

Without any prior input we expect that the mixing angle $\theta_\gamma$ to be $O(1)$ so that both $Z_{1,2}$ can now have 
substantial couplings to the dark sector fields carrying $Q_D\neq 0$ which may lead to important phenomenological implications.  The resulting mass-squared eigenvalues (always with 
$M_{Z_1}\geq M_{Z_2}$) are now given by 
\begin{equation}
\frac{2M_{Z_{1,2}}^2}{(g_Lc_wzv_R)^2}=2\lambda^2_{1,2}=1+\gamma^2(1+R)\pm \big[(1+\gamma^2)^2-2\gamma^2(1-\gamma^2)R+R^2\gamma^4\big]^{1/2}\,. 
\end{equation}
Here, the $\lambda_i$ (with $\lambda_1 \geq \lambda_2$) can be thought of as the masses of these new heavy neutral gauge bosons scaled in comparison to that of the conventional 
LRM expectation `reference' value for $M_{Z_R}$, \ie, $M_{Z_R}^2=(g_Lc_wzv_R)^2=M_0^2$.  It is to be noted that $M_0$ is itself a function of the parameter $\kappa$ and can 
vary significantly as the 
value of $\kappa$ changes; this dependence can be seen in the middle panel of Fig.~\ref{fig0}. Here we observe that $M_0$ diverges as $\kappa$ approaches its minimum value, 
$\simeq 0.55$, and that it grows linearly with $\kappa$ for larger values $\gsim 1$. The lower panel of this same Figure displays the $R,\gamma$ dependence of both $s_\gamma$ and 
$c_\gamma$; note that very substantial mixing occurs even for values of $\gamma$ below unity. While $s_\gamma$ grows linearly with $\gamma$ for small values, it rapidly asymptotes to 
unity; on the other hand, $c_\gamma$ falls like $1/\gamma$ for large values.

\begin{figure}
\centerline{\includegraphics[width=4.5in,angle=0]{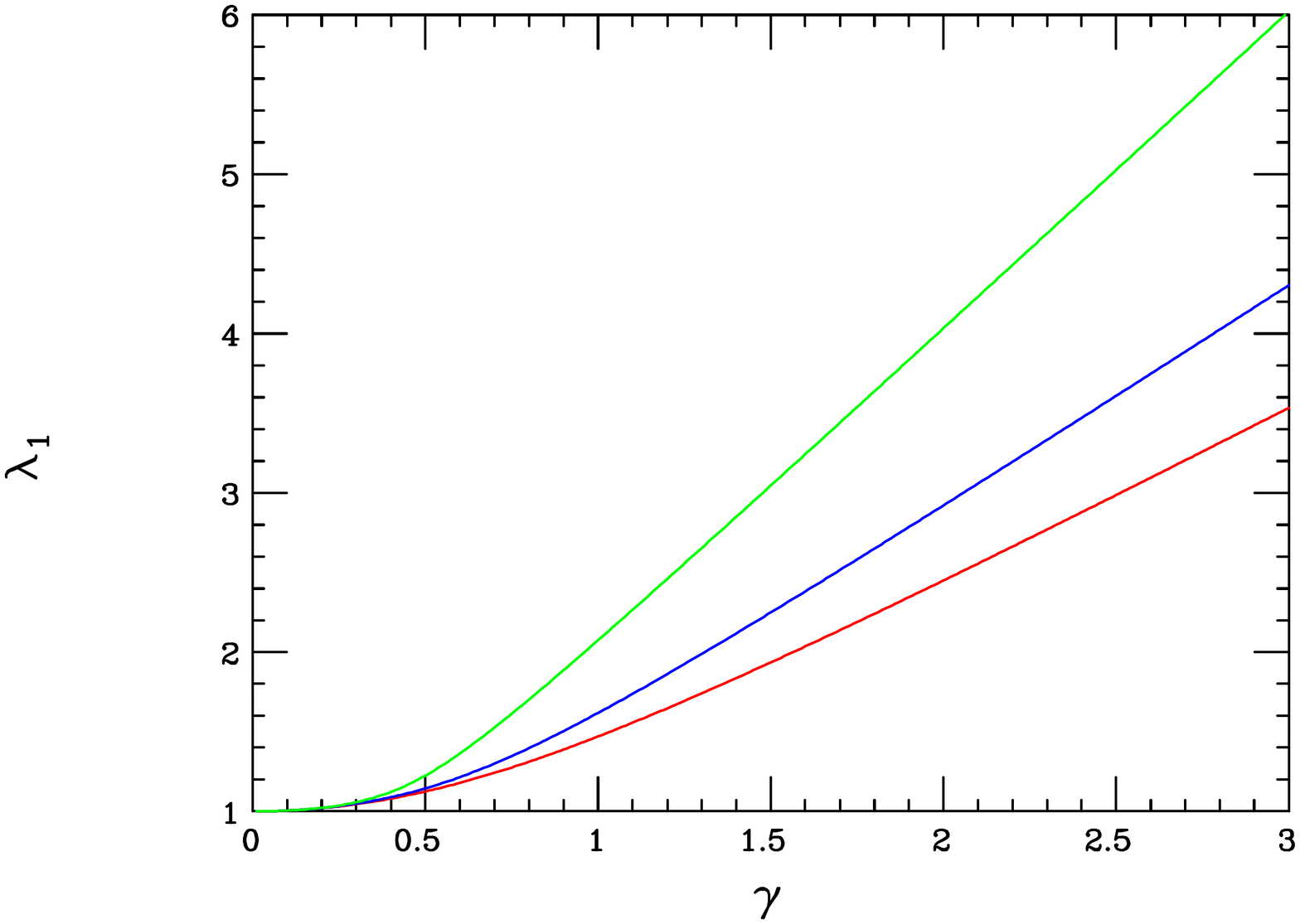}}
\vspace*{-2.0cm}
\centerline{\includegraphics[width=4.5in,angle=0]{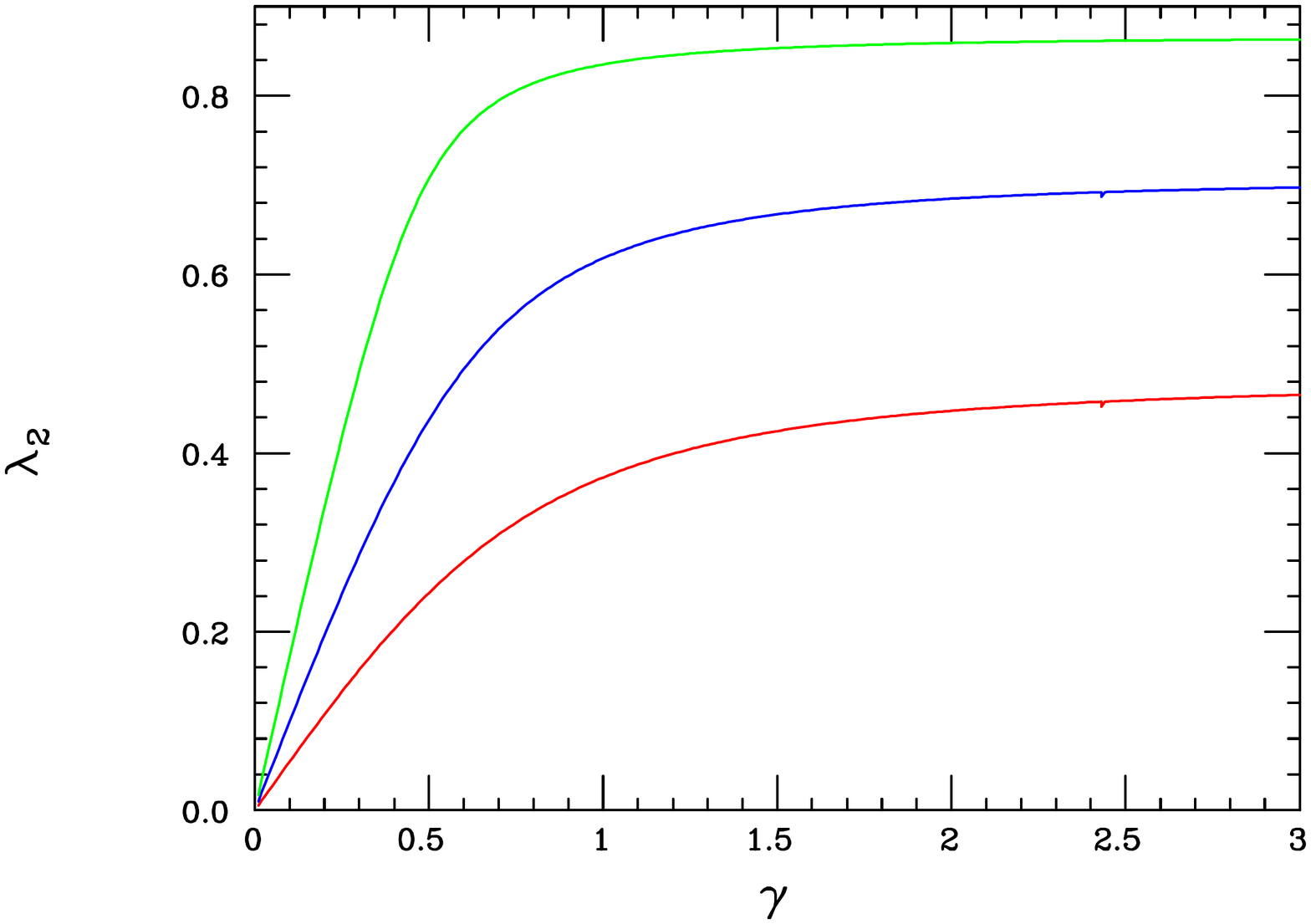}}
\vspace*{-2.0cm}
\centerline{\includegraphics[width=4.5in,angle=0]{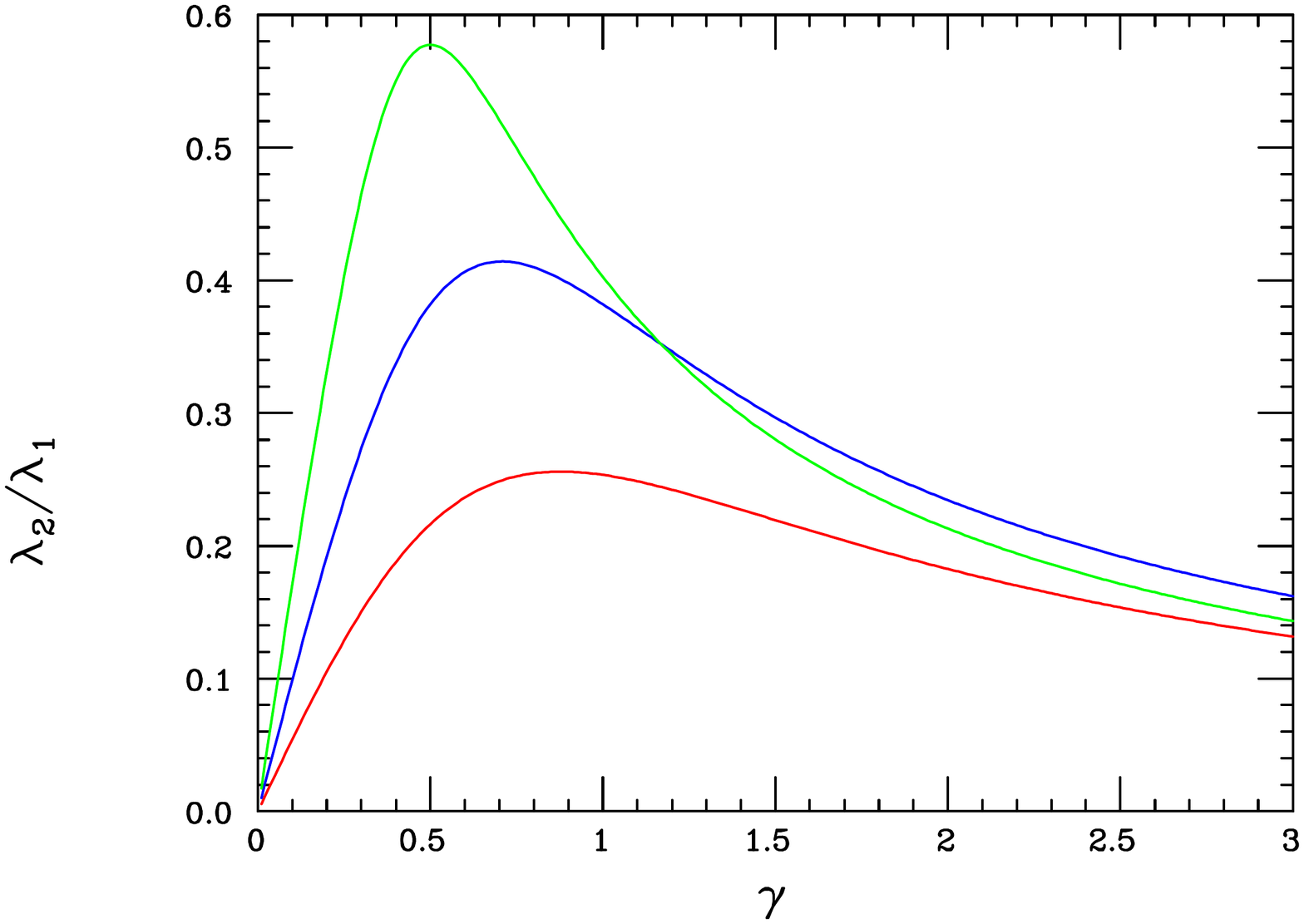}}
\vspace*{-1.30cm}
\caption{The scaled masses of the $Z_{1,2}$ gauge bosons, $M_{Z_i}=M_0 \lambda_i$, as described in the text, as functions of $\gamma$ for values of the vev ratio $R$=0.3 (bottom red), 
1 (middle blue) and 3 (top green), respectively. The top (middle) panel shows the result for $Z_{1(2)}$ while the lower panel show the corresponding gauge boson mass ratio. Note that all 
$\kappa$-dependence of these masses lies in the $M_0$ pre-factor. }
\label{fig1}
\end{figure}

Fig.~\ref{fig1} shows how these scaled $Z_i$ masses, the $\lambda_i$, vary as functions of the coupling ratio $\gamma$ for fixed values of the Higgs 
vev ratio, $R$. For large values of $\gamma$, $\lambda_1$ is found to grow asymptotically as $\sqrt{(1+R)} \gamma$. 
Note that $\lambda_2$ vanishes when $\gamma=0$, since the relevant gauge coupling then vanishes, and it then asymptotes at large values of $\gamma$ to $\sqrt{R/(1+R)}$. 
Here we see, \eg, that the $Z_2$ is always significantly lighter than $Z_1$ so it is likely to be much more kinematically accessible to collider searches (for fixed $M_0$) although both fields 
generally have qualitatively distinct couplings over much of the parameter space as we will find below. 

\begin{figure}[htbp]
\centerline{\includegraphics[width=5.0in,angle=0]{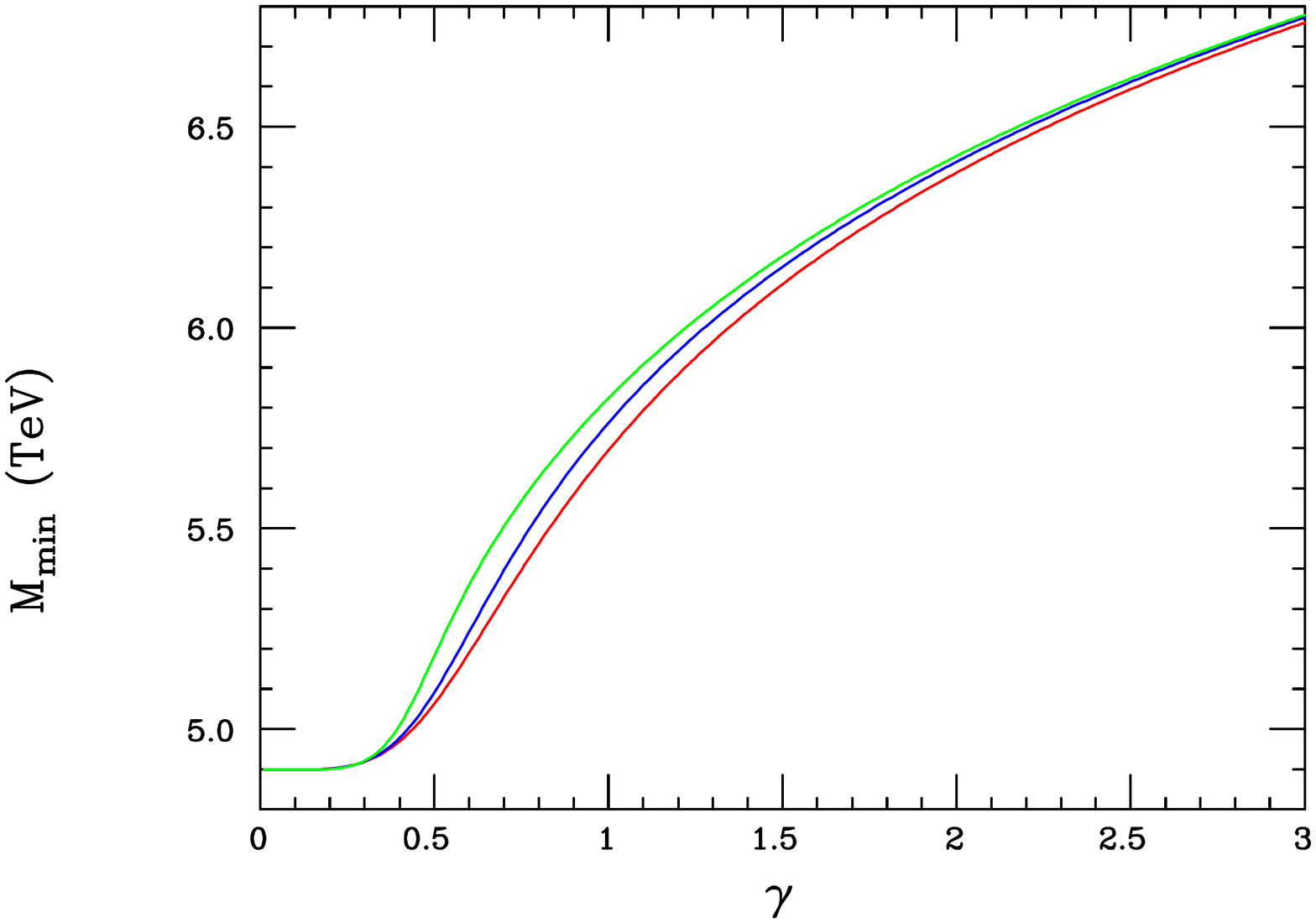}}
\vspace*{-2.0cm}
\centerline{\includegraphics[width=5.0in,angle=0]{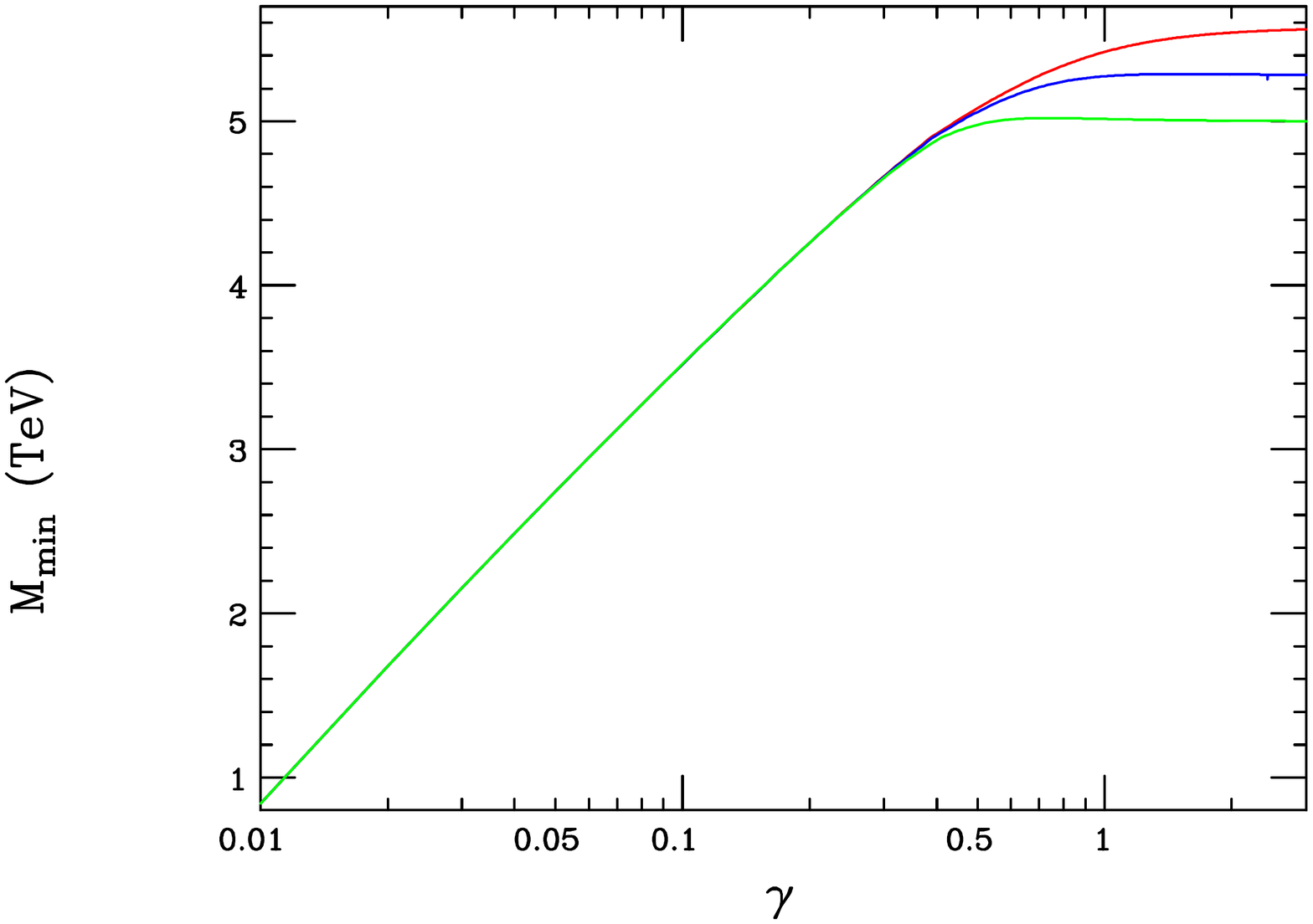}}
\vspace*{-1.50cm}
\caption{Approximate present lower bound on the (top) $Z_1$ and (bottom) $Z_2$ masses assuming only decays to SM fermion final states are kinematically accessible, as discussed in the 
text, as a function of the parameter $\gamma$ obtained by employing the dilepton search results from ATLAS\cite{ATLAS:2019erb}. The red (blue, green) curves correspond to 
$R=0.3(1,3)$, respectively, and, for demonstration purposes, all curves assume that $\kappa=1$ as well as  the applicability of the Narrow Width Approximation.}
\label{fig2}
\end{figure}

If $\kappa \simeq R \simeq 1$ and $\gamma$ is relatively small so that we are not too far from the LRM limit and only decays to the SM fermion final states are kinematically allowed, then, 
employing the results from the 13 TeV, 139 fb$^{-1}$  search by ATLAS\cite{ATLAS:2019erb}, we find that, \eg, the $Z_1$ is constrained from searches from the combined 
$e^+e^- + \mu^+\mu^-$ dilepton channel to lie above roughly $\simeq 4.9-5.1$ TeV. This constraint may increase by $\simeq 10-15\%$ or so as the 
LHC integrated luminosity is increased to 3 ab$^{-1}$\cite{Helsens:2019bfw} if no signal is found. Of course, if these various assumptions are significantly relaxed, the present search 
reach will extend over a significantly larger range of masses. For the $Z_1$, other regions of the parameter space can generally lead to stronger constraints than the one obtained in the LRM 
(always under the assumption that only decays to SM final states are kinematically allowed) as both the couplings to the SM quarks as well as the $Z_1$ leptonic branching fraction 
all increase with corresponding increases in values of $\gamma$. This result for the present $Z_1$ search limit, assuming the validity of the Narrow Width Approximation (NWA), is 
demonstrated in the upper panel of Fig.~\ref{fig2} where the choice $\kappa=1$ is maintained but both 
$R,\gamma$ are allowed to vary. If additional decay modes are present, clearly the $Z_1$'s branching fraction to SM leptons will diminish by, certainly at least, $O(1)$ factors which will 
degrade the search reach in this channel somewhat but this may be partially compensated for by the additional alternate search channels that now become available. 
Similarly, to the $Z_1$ example, the mass of the $Z_2$ is also constrained; however, in that case, as $\gamma \to 0$, the couplings of the $Z_2$ to SM states all vanish so that the 
bound then disappears. Of course, ifor larger values of $\gamma$, a respectable bound is obtained as the relevant couplings (initially) grow rapidly and 
this is shown in the lower panel 
of Fig.~\ref{fig2} under the same assumptions as were previously made for the $Z_1$. Since the $Z_2$ couplings saturate as $\gamma$ gets large, with $s_\gamma \to 1$ and 
$\gamma c_\gamma$ scaling approximately as $\sim (1+R)^{-1}$ independently of $\gamma$; here we see that the resulting bound flattens out in this parameter space region. Of course, 
once $\gamma$ becomes too large, depending upon the values of the other parameters, our reach estimate based on the NWA will fail as the $Z_i$'s will become too wide and thus the signal 
to background ratio under the resonance will drop significantly so that the limit obtained here will clearly overestimate the true bound by a potentially significant factor and a far more 
detailed analysis will then be required.

\begin{figure}[htbp]
\centerline{\includegraphics[width=5.0in,angle=0]{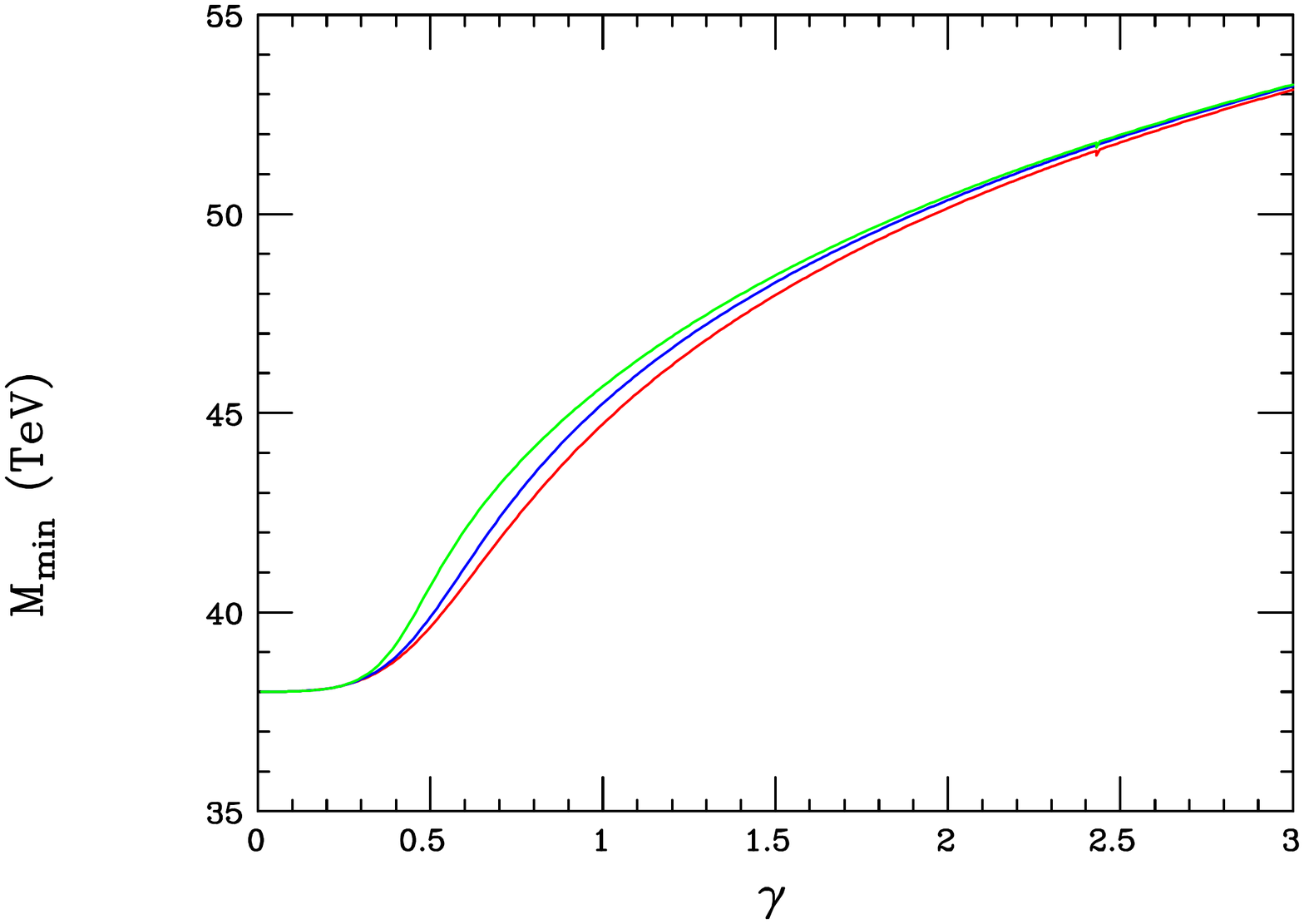}}
\vspace*{-2.0cm}
\centerline{\includegraphics[width=5.0in,angle=0]{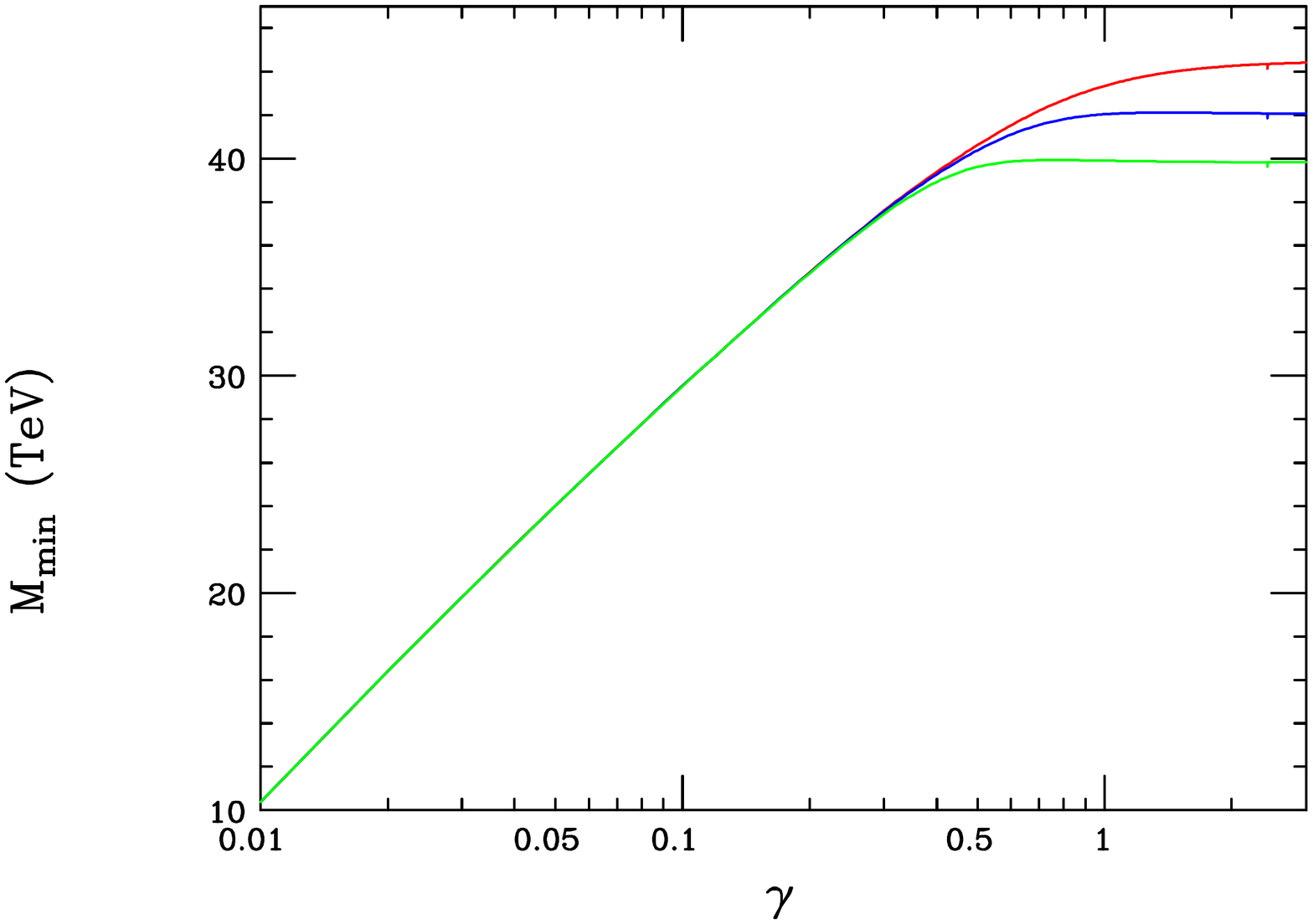}}
\vspace*{-1.50cm}
\caption{Same as the previous Figure, but now for the 100 TeV FCC-hh assuming an integrated luminosity of 30 ab$^{-1}$ following Ref.\cite{Helsens:2019bfw}.}
\label{fig3}
\end{figure}

It is also possible to extrapolate these results for the $Z_{1,2}$ mass reaches to the case of the 100 TeV FCC-hh by following the dilepton analysis as presented in Ref.\cite{Helsens:2019bfw}, 
here assuming an integrated luminosity of 30 ab$^{-1}$. The results of this analysis, with the same assumptions as in the case of the LHC are displayed in Fig.~\ref{fig3}, and unsurprisingly 
show the same overall qualitative behavior as was seen above although at a significant higher mass scale. The same words of caution with respect to the applicability of the NWA will apply 
in this case as above. 

Correspondingly, at this same mass scale of symmetry breaking, in the complex, non-hermitian sector, we find the relatively simple results for the gauge boson mass eigenvalues to be
\begin{equation}
M_{W_R}^2=\frac{1}{2}\kappa^2g_L^2v_R^2,~~~~M_{W_I}^2=\frac{1}{2}g_I^2v_R^2(1+2R)\,,
\end{equation} 
which, of course, cannot mix together as $Q(W^\pm,W_I^{(\dagger)})=\pm 1,0$ while $W_I$ also carries $|Q_D|=1$. The $W_R$ appearing here is just the usual one present in the 
LRM, except now for possible decay modes into heavy PM states and that its mass is no longer directly correlated in a simple manner with that of either of $Z_{1,2}$, for which many 
searches exist in multiple final states under various assumptions{\footnote {For a selection of such searches, see Refs.~\cite{wpsearch,ATLAS:2022jsi} }}. We note, however, that 
more indirectly, since all of the gauge boson masses are essentially 
determined by the values of $\kappa, \gamma, c_I/c_w$ and $R$, some correlations will exist especially in certain limits, \eg, at large values of $\gamma$ when $Z_1$ is mostly $Z_I$, 
$M_{W_I}$ is linearly proportional to $M_{Z_1}$. Roughly speaking, these lower bounds on the $W_R$ mass for $\kappa=1$ from these various LHC searches hover in the 
4.0-5.7 TeV range depending upon the search channel and will likely be improved upon somewhat by HL-LHC. 
Similarly, the mass of the $W_I$ is no longer directly correlated in a simple way with that of either $Z_{1,2}$, \eg, $M_{W_I}=c_I M_{Z_I}$, yet 
still must decay, if kinematically allowed, into a, $fF_{1,2}$, \ie, a SM+PM final state or, if this kinematically forbidden, into the $\bar ff+A_I$ final state as discussed in Ref.\cite{Rizzo:2022qan}. 
At the LHC, the $W_I$ can be made in pairs via $q\bar q$ annihilation via $s$-channel $Z_{1,2}$ exchange (plus $t$-channel $Q_{1,2}$ PM exchange), in association with $A_I$ (also via 
$t-,u$-channel $Q_i$ exchange), or in association with a $Q_{1,2}$ PM field in $gq$ fusion as was discussed in some detail in Ref.\cite{Rueter:2019wdf}. The 
situation here is slightly different, however, in that the $q=u$ channel for associated production is now also open. Since $W_I$ decay (to a very good approximation) necessarily involves 
PM fields, the search reaches for these states are much more model-dependent than are those for the other gauge bosons that we have so far discussed. Clearly, since $W_I$ production 
itself generally involves other heavy states at some level, the $W_I$ search reaches are clearly suppressed in comparison to those for the more well-studied $W_R$.

\begin{figure}
\centerline{\includegraphics[width=4.8in,angle=0]{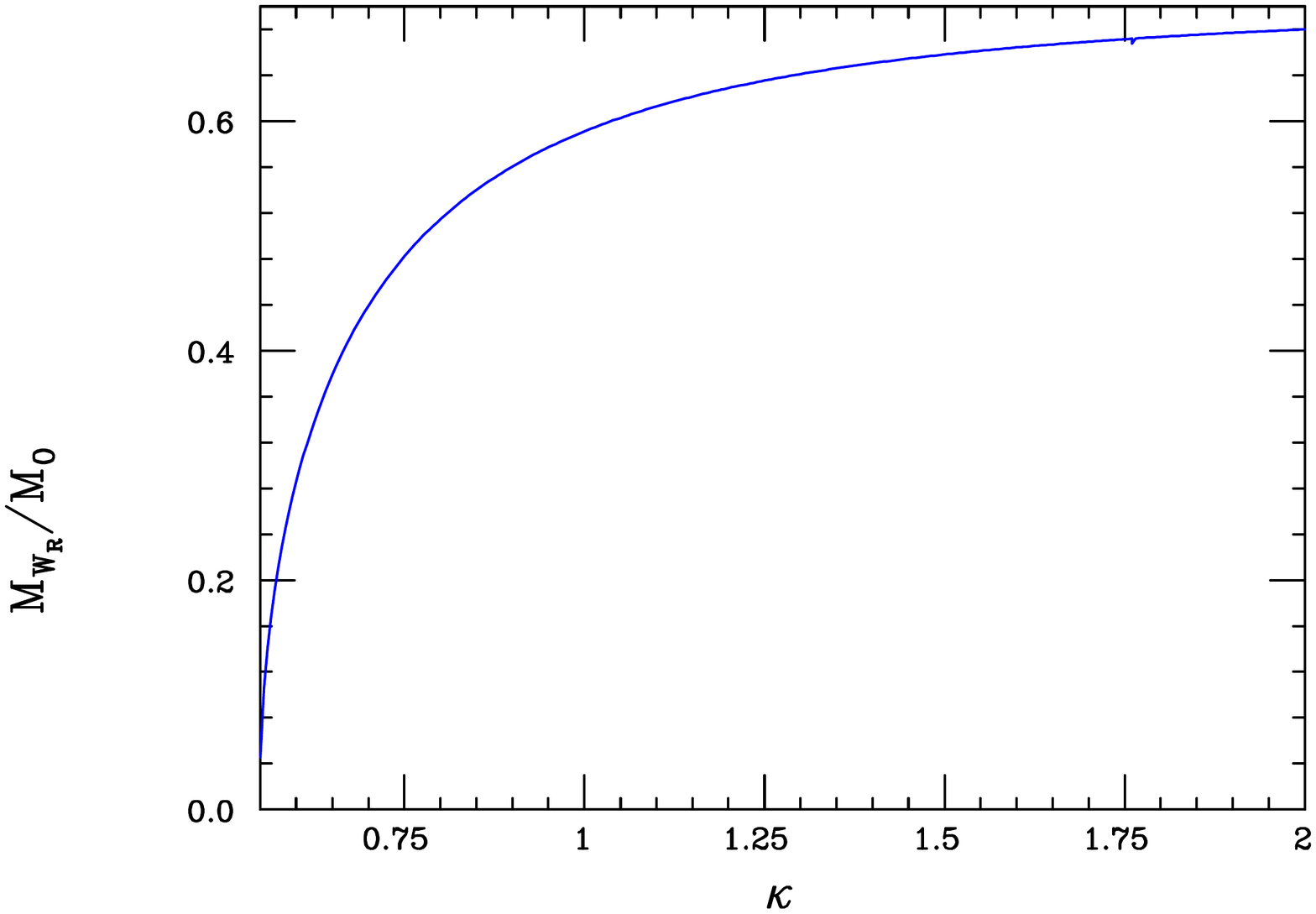}}
\vspace*{-2.0cm}
\centerline{\includegraphics[width=4.8in,angle=0]{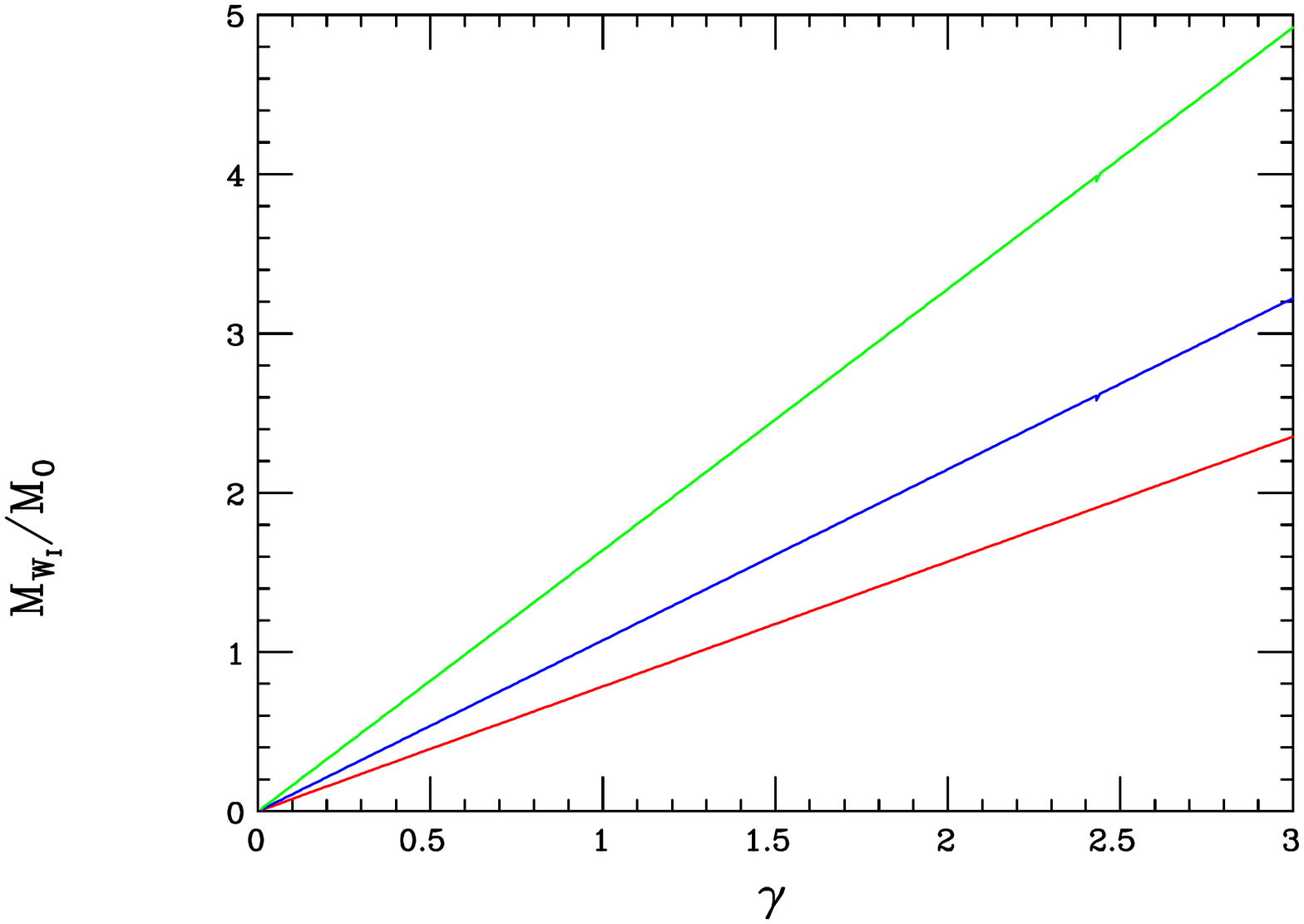}}
\vspace*{-1.30cm}
\caption{(Top) The $W_R$ mass, which is independent of the values of $\gamma,R$, in units of $M_0$ as a function of $\kappa$. (Bottom) The $W_I$ mass, which is $\kappa$-independent,  
in units if $M_0$ as a function of $\gamma$ for values of the vev ratio $R$=0.3 (bottom red), 1 (middle blue) and 3 (top green), respectively. Note that this value must be further rescaled by 
the ratio $c_I/c_w \leq 1.14$ which we have taken to be unity here for demonstration purposes.}
\label{fig4}
\end{figure}

To get an idea where these two non-hermitian gauge boson masses may lie relative to the those of the $Z_i$ discussed above, Fig.~\ref{fig4} shows both $M_{W_R}/M_0$ as a function of 
$\kappa$ (as it is independent of both $\gamma$ and $R$) and $M_{W_I}/M_0$ as a function of $\gamma$ (since it is independent of $\kappa$) for different values of $R$ {\it assuming} that 
an additional overall scaling factor of $c_I/c_w \lsim 1.14$ appearing in this mass ratio has been set to unity. Here we see that while both $W_R$ and $W_I$ generally lie somewhat close to the 
$Z_{1,2}$ in mass, it is difficult to  
make too many universal statements that might be useful, \eg, for resonance searches and/or model testing purposes at the LHC.  One obvious condition we observe is that the $W_R$ is 
{\it always} lighter than $Z_1$, being somewhat closer to the $Z_2$ in overall mass range. Indeed, for a respectable fraction of this parameter space the decay $Z_1\to W^+_RW^-_R$ 
is kinematically allowed. On the other hand, for much of the parameter space examined here, the $W_I$ is close to, but is always below, the $Z_1$ in mass even when the ratio $c_I/c_w$ 
takes on its maximum allowed value of $\simeq 1.14$. 

We next turn to the symmetry breaking which occurs at the electroweak scale, this time first examining the non-hermitian sector which is somewhat simpler as $W_I$ is unaffected by the 
relevant electroweak scale vevs of the $H_{1,4}$. The situation in the charged $W-W_R$ sector is essentially the same as in the LRM so we can be quite brief. 
Recalling the bi-doublet discussion above when these are projected back into the $2_L2_R$ subspace (\ie, the usual LRM subspace), one now generates a mass for the SM $W$, a shift 
in the $W_R$ mass, as well as a mixing between these two states as usual as can be seen from the $2\times 2$ mass matrix:
\begin{equation}
M^2_{WW_R}=\begin{pmatrix} M_W^2 & \beta_W M_W^2 \\ \beta_WM_W^2& M_{W_R}^2+\kappa^2M_W^2 \\ \end{pmatrix}\,,
\end{equation} 
where $M_{W_R}^2$ is as given above and 
\begin{equation}
M_W^2=\frac{1}{4}g_L^2(k_1^2+k_2^2+k_1'^2+k_2'^2), ~~~~~\beta_W=\kappa ~\frac{2(k_1k_2+k_1'k_2')}{k_1^2+k_2^2+k_1'^2+k_2'^2}\,.
\end{equation} 
This mass squared matrix can be diagonalized via a mixing angle (which we might expect to be $\sim 10^{-(3-4)}$) given in the notation above by
\begin{equation}
\tan 2\phi_W=\frac{-2\beta_W M_W^2}{M_{W_R}^2+(\kappa^2-1)M_W^2}\,,
\end{equation} 
to form the mass eigenstates $W_{1,2}$ given by $W=c_{\phi_W} W_1-s_{\phi_W} W_2$, \etc, and whose corresponding mass-squared eigenvalues are given by  
\begin{equation}
2M_{W_{1,2}}^2=M_{W_R}^2+(\kappa^2+1)M_W^2 \pm \big(M_{W_R}^4+2M_W^2M_{W_R}^2(\kappa^2-1)+M_W^4[4\beta_W^2+(\kappa^2-1)^2]\big)^{1/2}\,. 
\end{equation}
The resulting leading order fractional {\it downward} shift in the SM $W$ mass due to this mixing is then found to be very roughly of the same magnitude as the mixing angle, $\phi_W$, \ie, 
\begin{equation}
\frac{\delta M_W^2}{M_W^2} \simeq -\beta_W^2 \frac{M_W^2}{M_{W_R}^2} \,. 
\end{equation}

We next examine the corresponding symmetry breaking in the hermitian gauge boson sector at the electroweak scale where the situation is a bit more complex. Employing the SM relation 
$M_Z^2=M_W^2/c_w^2$ and recalling the $O(1)$ parameter combination employed above, $\Omega^2=\kappa^2-(1+\kappa^2)x$, in the now $Z-Z_1-Z_2$ basis the relevant $3\times 3$ 
mass squared matrix now becomes 
\begin{equation}
M^2_{Z,Z_1,Z_2}=\begin{pmatrix} M_Z^2 & -M_Z^2\Omega c_\gamma & M_Z^2\Omega s_\gamma \\ -M_Z^2\Omega c_\gamma& M_1^2+(M_Z\Omega c_\gamma)^2 & -M_Z^2\Omega^2 s_\gamma c_\gamma\\  M_Z^2\Omega s_\gamma & -M_Z^2\Omega^2s_\gamma c_\gamma &M_2^2+(M_Z\Omega s_\gamma)^2 \end{pmatrix}\,,
\end{equation} 
where $M_{1,2}^2$, $s_\gamma$, \etc, are all as defined above. To leading order in the small ratios $M_Z^2/M_{1,2}^2$, the most important effects that result from the $Z-Z_{1,2}$ mixings 
via the angles,  
\begin{equation}
\theta_{ZZ_{1,2}} \simeq \Omega M_Z^2 \Big( \frac{-c_\gamma}{M_1^2}, ~\frac{s_\gamma}{M_2^2} \Big) \,, 
\end{equation}
respectively, are to slightly reduce the SM $Z$ mass (but only by fractional factors $\sim 10^{-(3-4)}$ which are also the expected sizes of these mixing angles), \ie, 
\begin{equation}
\frac{\delta M_Z^2}{M_Z^2} \simeq -\Omega^2 M_Z^2 \Big(\frac{c_\gamma^2}{M_1^2} + \frac{s_\gamma^2}{M_2^2} \Big) = -\Omega^2 ~\frac{M_Z^2}{M_0^2}~\frac{1+R}{R} \,, 
\end{equation}
and to allow this (almost) SM $Z$ state to now pick up, at this mixing suppressed level, some of the couplings associated with both $Z_{R,I}$, \eg, a coupling to RH-neutrinos as well as 
to the set of dark sector fields which have $Q_D\neq 0$. It should be noted that over almost all of the model parameter space, one finds that $|\delta M_Z^2 |> |\delta M_W^2|$, employing the 
result obtained above and this may be of some interest given the recent $W$ boson mass measurement by CDF II\cite{CDF:2022hxs} due to the relative displacement of the two mass 
eigenstates induced via this mixing. 

The final stage of symmetry breaking at or above the electroweak scale arises from the effects of both the KM and the rather large set of all the possible $Q_D\neq 0$ vevs that may be 
non-zero in the 
various Higgs scalar representations we have introduced above; we'll deal with the KM effects first. In the hermitian sector, following the notation above and now accounting for the 
effects of the $\sim 10 $ TeV scale mass mixing discussed previously, the KM-induced interaction term in Eq.(9) above now appears in terms of the approximate mass eigenstates as  
\begin{equation}
{\cal L}^h_{int}({\rm {KM}})= \sigma ~\frac{g_Ls_w\kappa}{\Omega}~\big(Q_{em}-T_{3L}-T_{3R}\big)~ \big[ c_IA_I-s_I(c_\gamma Z_2+s_\gamma Z_1)\big]\,,
\end{equation}
with the further effects of the $Z-Z_{1,2}$ mass mixing at the electroweak scale being additionally suppressed by factors of order $\theta_{ZZ_{1,2}}$ that can be safely numerically neglected. 
Recalling the dimensionless 
parameters $\lambda_{1,2}$ from above,  we see that largest mass mixing term generated by this interaction arises unsurprisingly from the $T_{3I}=T_{3R}=1, T_{3L}=0$ 
vev, $v_R$, resulting in the induced mass mixing of $A_I$ with the $Z_{1,2}$ via both new diagonal and off-diagonal terms given by 
\begin{equation}
\theta_{A_IZ_{1,2}} \simeq -\frac{\sigma t_wc_I}{\kappa}~ \Big( \frac{c_\gamma+\gamma s_\gamma}{\lambda^2_1}, ~\frac{\gamma c_\gamma-s_\gamma}{\lambda^2_2} \Big) \,, 
\end{equation}
which are expected to be roughly $\sim 10^{-4}$ or so, such that to leading order in the small parameters, one essentially finds the corresponding shifts in the fields  
\begin{equation}
A_I\to A_I+\sum_i~\theta_{A_IZ_i}Z_i, ~~~~Z_i\to Z_i-\theta_{A_IZ_i}A_I\,,
\end{equation}
result in the diagonalization of the perturbed mass squared matrix. 
This implies that both $Z_{1,2}$ pick up some KM-suppressed interactions to $T_{3I}=0, Q_D\neq 0$ states that they might otherwise not have coupled to, while the $A_I$ correspondingly 
picks up  KM-suppressed couplings to the SM fields with non-zero values of $T_{3(L,R,I)}$ and/or $Q_{em}$ which all have $Q_D=0$. Combining the $\theta_{A_IZ_i}$-induced couplings 
here with those in
Eq.(40) (and recalling that the $A_I$ direct coupling to $Q_D$ is already present at leading order), after some algebra we now find that the {\it total} KM-induced coupling for $A_I$ at this stage of 
symmetry breaking to SM/LRM states is explicitly given by (and recalling from above that $\kappa^2 > t_w^2$) 
\begin{equation}
\sigma g_Yc_I \Big(1-\frac{t_w^2}{\kappa^2}\Big)^{-1/2}~ \Big[ (\alpha-1)T_{3R}+\Big(1- \frac{\alpha t_w^2}{\kappa^2}\Big)\frac{Y}{2} +\beta \gamma T_{3I}\Big]\,,
\end{equation}
where $\gamma$ is given above, $g_Y$ is the usual SM hypercharge coupling and, in terms of the previously defined parameters, one finds that the coefficients $\alpha,\beta$ are given by 
\begin{equation}
\alpha = \frac{c_\gamma^2+\gamma s_\gamma c_\gamma}{\lambda^2_1}-\frac{\gamma s_\gamma c_\gamma -s_\gamma^2}{\lambda^2_2},~~~~\beta = \frac{\gamma s_\gamma^2+
s_\gamma c_\gamma}{\lambda^2_1}+\frac{\gamma c_\gamma^2 -s_\gamma c_\gamma}{\lambda^2_2}\,.
\end{equation}
Note that in the pure LRM limit, \ie, $\gamma,s_\gamma, \beta \to 0$ so that also $c_\gamma, \alpha \to 1$, the DP coupling is easily seen to be only to the SM hypercharge at this stage of 
symmetry breaking as it would be in the familiar $U(1)_D$ DP model.  In this same limit we would then easily identify the usual $\epsilon$ parameter of the $U(1)_D$ model to be given by 
\begin{equation}
\epsilon=\sigma c_w c_I \Big(1-t_w^2/\kappa^2 \Big)^{1/2}\,. 
\end{equation}
Interestingly, using the definitions above, after some lengthy algebra one finds that the relations $\alpha=1, \beta=0$ are {\it always} satisfied so that the $A_I$ in this setup indeed only has 
KM-induced couplings to the $Q_D=0$ sector via the SM hypercharge as in the usual $U(1)_D$ model.

The mass of $A_I$, \ie, $M_{A_I}$ , which we've not yet discussed in any detail as, before any potential mixing effects, it arises solely from the $Q_D\neq 0$ vevs, is found not to be shifted 
to leading order in the small parameters by KM but it is possible that the quadratic terms of order $\sigma^2 M_0^2$ can potentially be present and could be numerically significant in some 
regions of the parameter space as we expect $\sigma \sim \epsilon \sim 10^{-(3-4)}$ and $M_0$ is relatively quite large, at least several TeV.  In order to address this potential problem, we must 
return to the analysis above and re-examine the full $3 \times 3$, $Z_R-Z_I-A_I$ mass-squared matrix including these new terms that are now generated by KM:
\begin{equation}
M_0^2~\begin{pmatrix} 1 & \gamma & -q\\  \gamma & \gamma^2(1+R) & -\gamma q\\  -q & -\gamma q & \tilde M_{A_I}^2 +q^2 \\ \end{pmatrix}, ~~~~~ q=\frac{\sigma t_w c_I}{\kappa},~~~~~\tilde M_{A_I}=M_{A_I}/M_{0}\,.
\end{equation} 
Combining all of the contributions to the squared $A_I$ mass, one finds that, fortunately, the the terms which are quadratic in $\sigma$ completely cancel so that the DP mass still only 
arises from the vevs of the $Q_D\neq 0$ scalars that we will discuss later below.  This is a generalization of the well-known result that occurs in the simple $U(1)_D$ scenario.

Next, we consider whether or not $A_I$ and $Z$ will correspondingly 
mix in the familiar manner via the electroweak scale, $B-L=0$, bi-doublet vevs from $H_{1,4}$ that were discussed previously. We recall 
that the initial KM-induced $A_I$ coupling to the SM fields in Eq.(40) is proportional to $B-L=Q -T_{3L}-T_{3R}$ so this coupling would vanish completely for these 
representations. However, the $A_I-Z_i$ mass mixing above was seen to alter this situation as $A_I$ now in general couples instead only to $Y/2$ implying a non-vanishing contribution 
from the bi-doublets which appears in the conventional manner. The relevant $Z-A_I$, $2\times 2$ part of the gauge boson mass squared matrix can be written at this level of approximation, 
now employing the conventional $\epsilon$ notation, as 
\begin{equation}
\begin{pmatrix} M_Z^2 & -\epsilon t_w M_Z^2 \\ -\epsilon t_w M_Z^2 & M_{A_I}^2+\epsilon^2 t_w^2 M_Z^2\\ \end{pmatrix}\,,
\end{equation} 
which can be diagonalized as usual by $Z\to Z+\epsilon t_w A_I$, etc. After diagonalization, the mass of $A_I$ is unaltered but it now couples as $\simeq \epsilon eQ$ in the limit when 
$M_{A_I}^2/M_Z^2 << 1$, appearing as the conventional DP as far as the KM-suppressed couplings are concerned.

By way of contrast, the non-TeV but now {\it electroweak} scale vev KM-induced $A_I-Z_i$ mixing is also found to be non-zero but it is significantly smaller than that obtained in the 
discussion above for KM-induced mixing with the $Z$ by factors of order $\sim M_Z^2/M_{1,2}^2 <10^{-(3-4)}$ and so can be safely neglected in what follows.

Lastly, we must turn our attention to the set of the many possible $Q_D\neq 0$ vevs, here denoted collectively as $w_i$, that can occur at scales $\lsim 1$ GeV and generally also have 
other additional quantum numbers, \eg, $T_{3(L,R,I)}$, associated with them depending upon which Higgs scalar representation of the many encountered above that they may come 
from. These will not only generate a mass (before any possible mass or kinetic mixing effects might be included) for $A_I$, \ie, 
\begin{equation}
M_{A_I}^2=g_D^2~\sum_i ~Q_{D_i}^2 w_i^2\,,
\end{equation}
but will also induce a small gauge boson mass mixing with the other neutral states (including now with the combination $W_I+W_I^\dagger$) as well as generating (relatively) tiny Majorana 
mass terms for some subset of the neutral fermions as was discussed above. The 
largest of the resulting hermitian gauge boson mixings will be induced between the $Z$ and $A_I$ as the corresponding mixing with $Z_{1,2}$ will be further suppressed by factors of order 
$\sim M_Z^2/M_{1,2}^2$, and so vevs with both $T_{3L},Q_D\neq 0$ will be the most relevant.  This precludes the $H_i$ as well as $\Delta_R,\tilde \Delta_R$ and $X_R$ 
from playing any important role in generating these mixings{\footnote {This will of course {\it not} be the case for the $A_I$ mass itself.}}. Thus the $Q_D\neq 0$ vevs of the 
three remaining Higgs scalars, $\Delta_L,\tilde \Delta_L$ and $X_L$, will be the main subject of our attention and the resulting $Z-A_I$ induced mixing angle can generically be written as 
\begin{equation}
\phi_{ZA_I}\simeq \frac{g_L}{g_Dc_w}~ \Big[\frac{\sum_i~Q_{D_i}T_{3L_i}w_i^2}{\sum_i~Q_{D_i}^2w_i^2}\Big]~\frac{M_{A_I}^2}{M_Z^2}\,, 
\end{equation}
where we expect the pre-factor in front of the mass squared ratio to be roughly $\lsim O(1)$. This implies a not too uncommon additional coupling of the $A_I$ to SM fields in a $Z$-like 
manner, \ie, proportional to $-\phi_{ZA_I}~\frac{g_L}{c_w}~(T_{3L}-xQ)$.  Here, in principle, the $A_I$ mass also experiences a tiny fractional shift, $\delta M_{A_I}^2/M_{A_I}^2$, due to 
this mixing as well from the KM-induced couplings to the $Z_{SM,1,2}$ associated currents so that {\it all} of the $Q_D\neq 0$ vevs may now contribute; however, these terms are all found to be
 suppressed by appropriate factors of order $M_{A_I}^2/M_{Z_{L,1,2}}^2$ and so can be safely ignored.

As has been noted several times, the $A_I$ is also found to have a somewhat unusual induced mixing with the $Q_D\neq 0$ hermitian combination $W_I+W_I^\dagger$ of states 
arising from the Higgs representations 
which are $SU(2)_I$ non-singlets. In particular, the largest contributions to this mixing will arise, due to the action of the raising and lowering operators, from the product of a $Q_D=0$ and 
a $Q_D\neq 0$ vev from representations wherein the 
largest $Q_D=0$ vevs reside, \ie, the $SU(2)_I$ doublets $H_{2,3}$ (with vevs $\Lambda,\Lambda'$, respectively) and the $SU(2)_I$ triplet $\Delta_R$ (with vev $v_R$). Let us denote 
the corresponding small, $Q_D=-1$,  $\lsim 1$ GeV vevs in these representations by $\lambda,\lambda'$ and $v'_R$, respectively, as above. Then the $(W_I+W_I^\dagger)-A_I$ mixing 
angle is found to given by 
\begin{equation}
\phi_{W_IA_I}\simeq -s_I~\frac{2v_Rv'_R+(\Lambda \lambda+\Lambda' \lambda')}{2v_R^2(1+2R)}\,, 
\end{equation}
which is again found to be roughly of order $\sim 10^{-4}$.  This implies that $A_I$ picks up a new, $Q_D$-changing coupling to the $SU(2)_I$ isospin raising and lowering operators of the form 
\begin{equation}
g_I \phi_{W_IA_I}~(T^+_I+T_I^-)~A_I\,.
\end{equation}
When acting on the $f-F_1-F_2$ fermions, this structure produces the effective interaction  
\begin{equation}
g_I\phi_{W_IA_I}~\Big[\bar f \gamma_\mu P_L\big(F_1c_L-F_2s_L\big)+\bar f \gamma_\mu P_R\big(F_2c_R+F_1s_R\big)\Big] A^\mu_I +\rm{h.c.}\,,
\end{equation}
which augments those couplings of a similar $Q_D$-violating nature already appearing above in Eq.(26) due to SM-PM fermion mixing effects. By adding these two results, we see that to 
leading order in the small vev ratios, the sum of both the contributions to the effective $f-F_1-A_I$ coupling can now be written as 
\begin{equation}
-g_D\bar f \gamma_\mu P_L F_1~\Big[\frac{\lambda}{\Lambda} + \frac{2v_Rv'_R+(\Lambda \lambda+\Lambda' \lambda')}{2v_R^2+\Lambda^2+\Lambda'^2}\Big]~A^\mu_I+\rm{h.c.}\,, 
\end{equation}
while that for $F_2$ is given in the same approximation by the same expression with the replacements $P_L \to  P_R$ and $(\Lambda,\lambda)\to (\Lambda',\lambda')$. Couplings to 
the opposite helicities 
are found to be suppressed for both $F_i$ states by factors of order $s_{L,R}\sim (v,v')/(\Lambda,\Lambda') \sim 10^{-2}$.  As noted previously, this result is qualitatively similar to that found 
in Ref.~\cite{Rueter:2019wdf} in a somewhat different (and simpler) context. Like in that case, here we also see that the contribution to the $F_{1,2}\to fA_I$ decay 
amplitude due to the longitudinal component of the $A_I$ polarization is enhanced by a factor of $m_{1,2}/M_{A_I}>>1$ which offsets the suppression due to the small overall mixing angle 
factor, $\phi_{W_IA_I}$ or, more explicitly, by the set of small vev ratios appearing in the expression above.

\begin{figure}
\centerline{\includegraphics[width=4.3in,angle=0]{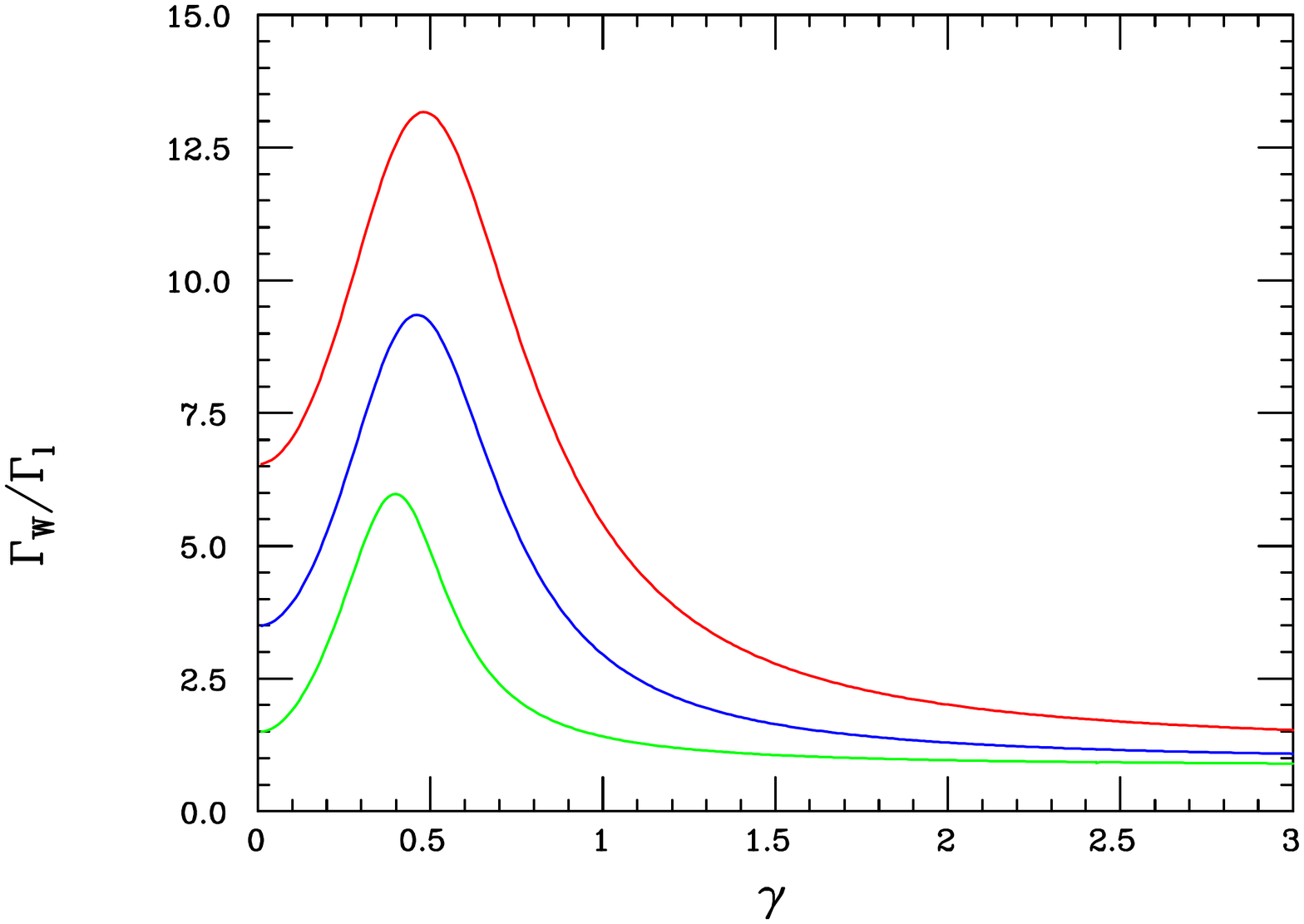}}
\vspace*{-2.0cm}
\centerline{\includegraphics[width=4.3in,angle=0]{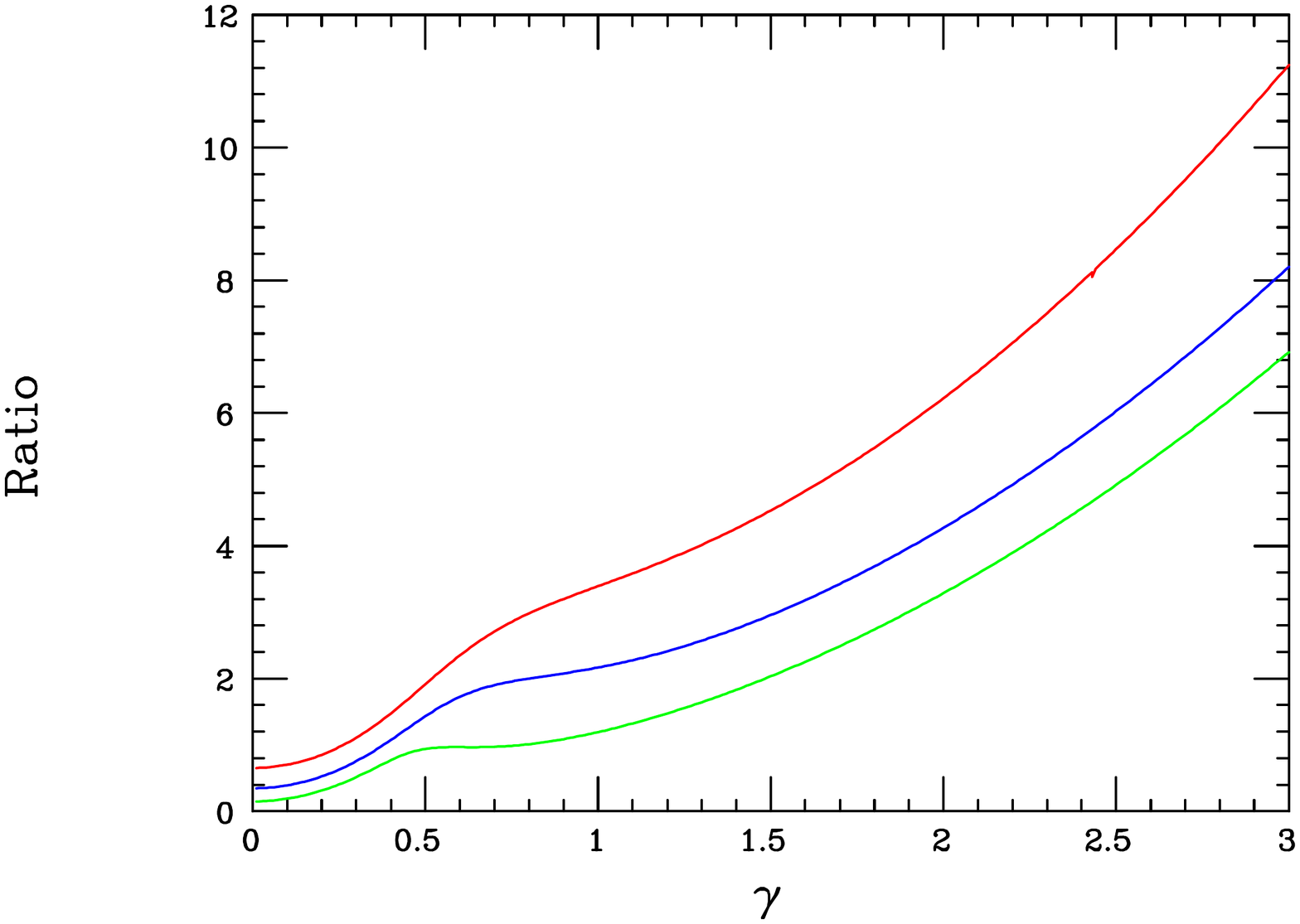}}
\vspace*{-2.0cm}
\centerline{\includegraphics[width=4.3in,angle=0]{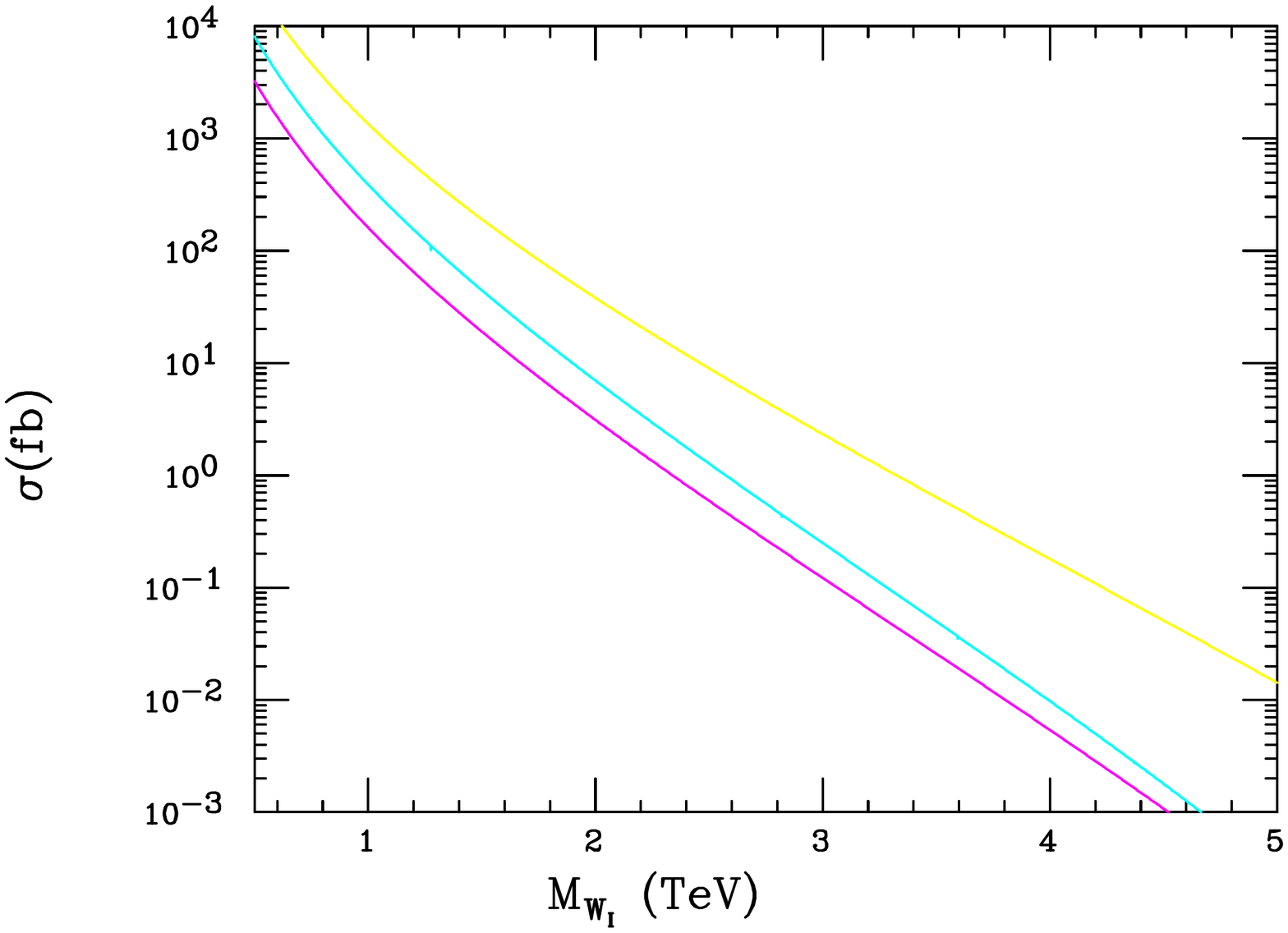}}
\vspace*{-1.30cm}
\caption{(Top) The ratio of the $W_I^{(\dagger)}A_I$ to dilepton partial widths of the $Z_1$ as a function of $\gamma$ for values of the vev ratio $R$=0.3 (top red), 1 (middle blue) and 
3 (bottom green), respectively, assuming $f{\cal R}^2=1$. (Middle) The ratio of the resonant $Z_1\to W_I^{(\dagger )}A_I$ production cross section at the 13 TeV LHC to that for dileptons in 
the LRM with the three curves labeled as in the panel above and here assuming that $f{\cal R}^2=0.1$. (Bottom) Resonant $Z_1 \to W_I^{(\dagger)}A_I$ induced production cross section as a 
function of $M_{W_I}$ assuming that $R=\gamma=1$ (bottom magenta), $R=0.3, \gamma=2$ (middle cyan) and $R=0.3,\gamma=3$ (top yellow) as an extreme example, together with 
$f{\cal R}^2=0.1$. Note that the values $\kappa=c_I/c_w=1$ have been assumed in all of the panels above.}
\label{fig6}
\end{figure}

A direct application of this analysis arising from the $(W_I+W_I^\dagger)A_I$ mixing-induced coupling is the process $Z_i \to W_I^{(\dagger)}A_I$ which occurs via 
the $Z_IW_IW_I^\dagger$ 
non-abelian trilinear interaction at order $g_I$. This is an $s$-channel, resonance-enhanced version of a previously examined process\cite{Rueter:2019wdf} which in that 
case instead occured via $t-$ (or $u-$)channel $F$-exchanges so that in the present case the $W_I$ (and its decay products) would appear more centrally in the detector. As discussed 
above, the large mass ratio $M_{Z_i}/M_{A_I}$ appearing in the amplitude due to the dominance of the $A_I$ longitudinal polarization offsets the small value of $\phi_{W_IA_I}$. Note that 
given the scaled $Z_i$ and $W_I$ masses shown in the Figures above, for most of the parameter space only the resonant decay with the $Z_1$ initial 
state will be kinematically allowed on-shell. To estimate the cross section for this process, we need several distinct pieces of information, \eg, the fraction of the $A_I$ mass resulting from the vev 
$v_R'$, \ie,  $f=(g_Dv_R')^2/M_{A_I}^2 < 1$. Then, we need to account for the various multiple vev ratios that enter into the $\phi_{W_IA_I}$ mixing angle 
expression as well as the gauge boson masses themselves; to this end we define the $O(1)$ ratio 
\begin{equation}
{\cal R}=(1+2R)^{-1}~\Big[1+ \frac{\Lambda \lambda+\Lambda' \lambda'}{2v_Rv_R'} \Big] \,,
\end{equation}
which we see equals unity when $\lambda/\Lambda=\lambda'/\Lambda'=v_R'/v_R$, but can be either greater or less than one. Next, to obtain the NWA estimate for the desired cross section, we 
need to determine the ratio of the $Z_1$ partial widths for the $W_I^{(\dagger)}A_I$ final state to that for dileptons, $\Gamma_W/\Gamma_\ell$, which we can write as 
\begin{equation}
\frac{\Gamma_W}{\Gamma_\ell}=f {\cal R}^2~\frac{s_\gamma^2 \lambda_1^2}{R_\ell^2+L_\ell^2}~P(M_{W_I}^2/M_{Z_1}^2)\,,
\end{equation}
where, $s_\gamma, \lambda_1$ are defined above, $R_\ell,L_\ell$ are the leptonic chiral couplings of the $Z_1$ which depend upon the parameters $R,\gamma$ and $\kappa$, and 
$P(y)$ is a kinematic function arising from the product of the squared matrix element with the relevant phase space:
\begin{equation}
P(y)=(1-y)^3 ~\Big[ 2+\frac{(1+y)^2}{4y}\Big]\,.
\end{equation}
The top panel of Fig.~\ref{fig6} shows this partial width ratio as a function of $\gamma$ for specific values of $R$ and we see that it is generally $O(1)$ or larger. In the middle panel, we 
take the ratio of the resonant $Z_1\to W_I^{(\dagger )}A_I$ production cross section at the 13 TeV LHC to that for dileptons as given in the LRM for reference here assuming that 
no other additional new $Z_1$ decay modes exist for simplicity (and thus avoiding a reduced branching fraction) as well as $\kappa=c_I/c_w=1$ and, to be a bit conservative, 
we will also take $f{\cal R}^2=0.1$ for purposes of demonstration. Finally, combining these results, we 
can obtain the corresponding resonant $(W_I+W_I^\dagger)A_I$ production cross section for specific values of ($\gamma,R$) as a function of the mass of the $W_I$ as is shown for  
three sample values of these pairs of parameters in the lower panel of Fig.~\ref{fig6}, again with the assumption that $f{\cal R}^2=0.1$.  Note, in particular, the very large result obtained 
in the case of $R=\gamma=3$; this is due not only to the enhanced $Z_1$ couplings one finds for these parameter choices, but also the fact that, for a fixed value of $M_{Z_1}$, a 
larger value of $M_{W_I}$ is obtained and here we are displaying the cross section 
as a function of this variable, {\it not} $M_{Z_1}$. These results compare quite favorably with those shown in the top panel of Fig.(13) in Ref.\cite{Rueter:2019wdf} for the non-resonant process 
mediated by quark-like PM (to which this new resonant contribution would be added) which was obtained under somewhat different assumptions.

Clearly, with such a rather 
complex setup with multiple moving parts in the gauge, fermion and scalar sectors, many more interesting processes which can be probed at colliders will arise; we plan to  
consider these and other potential signatures in our later work.

\section{Discussion and Conclusions}

The kinetic mixing portal model 
allows for the possibility of thermal dark matter in the sub-GeV mass range owing to the existence of a similarly light gauge boson mediator which has naturally 
suppressed couplings to the fields of the Standard Model. The success of this KM scenario for generating the interactions of SM fields with DM rests upon the existence of a new set of portal 
matter fields, at least some of which may naturally lie at the $\sim$ TeV scale\cite{Rizzo:2022qan}, which carry both SM and dark charges thus allowing for the generation of the link between 
the ordinary and dark gauge fields at the 1-loop level via vacuum polarization-like graphs. In the simplest abelian realization of this possibility, the dark gauge group, $G_{Dark}$, is just 
the $U(1)_D$ associated with the dark photon; SM fields are all neutral under $U(1)_D$, \ie, they have $Q_D=0$ and so do not couple directly to the DP except via KM.  However, the 
existence of PM indicates 
that a larger gauge structure of some kind for $G_{Dark}$ is likely present and one may then ask how $G_{SM}$ and this enlarged $G_{Dark}$ might fit together into a more unified 
framework.  Clearly, at least a partial answer to this question can be found through an understanding of the detailed nature and possible properties of the various scalar and/or fermionic PM 
fields themselves and an examination of other impacts that their existence might have beyond their essential role in the generation of KM. 

In past work, we have begun an examination of the interplay of PM and a simple non-abelian version of $G_{Dark}$ together with the SM following both complementary bottom-up and top-down 
approaches in an effort to gain insight into these and related issues given a minimal set of model building requirements (which included a finite and calculable value for the usual KM 
parameter, $\epsilon$, \cite{Rizzo:2018vlb,Rueter:2019wdf,Rueter:2020qhf,Wojcik:2020wgm,Rizzo:2021lob,Rizzo:2022qan,Wojcik:2022rtk,Rizzo:2022jti,Rizzo:2022lpm}.  Amongst the 
findings from this set of analyses are that ($i$) it is likely for at least some of the SM and PM fields lie in common representations of $G_{Dark}$, the simplest example of which, 
consistent with our constraints and the one employed here, being the SM-like $SU(2)_I\times U(1)_{Y_I}$ setup which breaks down to $U(1)_D$ at a mass scale essentially the same as 
that at which the 
PM fields acquire their masses. Again, paralleling the SM, this is achieved by having the PM masses generated by Yukawa couplings to dark Higgs fields whose vevs are also responsible for 
the breaking of $G_{Dark}$. It was also found that  ($ii$) it is possible to relate phenomenological issues in both the visible and dark sectors, \eg, the magnitude of a possible upward shift 
in the mass of $W$ relative to SM expectations, as measured by CDF II \cite{CDF:2022hxs}, can be related to the mass of the DP while also satisfying the other model constraints. ($iii$) In a 
top-down study based upon the assumption of a unification of $G_{SM}\times G_{Dark}$ in a single, though hardly unique, $SU(N)$ gauge group\cite{Rizzo:2022lpm}, it was found that all of 
our model building constraints could not be satisfied when both $G_{SM}$ and $G_{Dark}$ take their `minimal' forms. ($iv$) It was shown that there are some possible model building gains 
to be made when addressing various experimental puzzles by also extending $G_{SM}$ beyond the usual $3_c2_L1_Y$ while also simultaneously considering a non-abelian $G_{Dark}$ as 
was done earlier\cite{Wojcik:2020wgm} to relate the dark sector and the KM mechanism with the flavor/mixing problem. By employing a simpler, single generation version of this same model, in 
this paper we have begun to examine the possible relationship between the masses of the portal matter fields and the masses of the right-handed neutrino as well as the new spin-1 fields 
associated with both its visible and dark extended gauge sectors when the symmetries of the SM are replaced by those of the Pati-Salam/Left-Right Symmetric Model, 
\ie, $G=4_c2_L2_R2_I1_{Y_I}$ or, more simply below the color breaking scale of $M_c\gsim 10^6$ TeV which concerns us here, just $G_{eff}=3_c2_L2_R2_I1_{Y_I}1_{B-L}$.

Amongst the many immediate implications of and results obtained from this setup that we've examined above are that ($a$) it is $1_{Y_I}1_{B-L}$ that undergo abelian kinetic mixing at the 
$\sim$ few TeV scale;  ($b$) Left-Right symmetry plus 
anomaly cancellation requires the set of fermionic PM fields to transform as a complete vector-like family under both the SM/LRM as well as the $U(1)_D$ symmetries and this also leads to a 
finite and calculable value for KM strength parameter $\epsilon$ of the desired magnitude, $\sim 10^{-(3-4)}$. ($c$) All of the usual chiral SM/LRM fermion fields (which still carry $Q_D=0$) lie 
in doublets of $SU(2)_I$ together with a corresponding PM field (which has $Q_D=-1$) with which they share their QCD and electroweak quantum numbers.  At the TeV scale these two sets 
of fields are connected via the exchange of the neutral, non-hermitian gauge bosons of $SU(2)_I$, $W_I^{(\dagger)}$;  however, the DP also couples these two sets of fields at the 
sub-GeV scale but in a suppressed manner yielding the dominant PM decay path. ($d$) As usual, at low energies the DP couples diagonally to the SM via KM as $\simeq e\epsilon Q_{em}$ 
and, as occurs frequently in many setups, also proportional to the SM $Z$ couplings via mass mixing through a small mixing angle of the same order as $\epsilon$. ($e$) If the standard 
RH-triplet Higgs fields are employed to break the 
LR symmetry and generate a heavy Majorana see-saw mass for the RH-neutrino via a $|B-L|=2$ vev, since these RH-triplet Higgs are {\it additionally} required to be $SU(2)_I$ triplets, they will 
necessarily also lead to the breaking of $2_I1_{Y_I}\to 1_D$ at the same mass scale. The extra Higgs scalars generating the Dirac masses of the charged PM fermions will also contribute 
to this same symmetry breaking. ($f$) The same bi-triplet Higgs representations also contain vevs carrying both $|Q_D|=1,2$, the later of which contributes to a tiny splitting in the masses 
of each of the two heavy neutral  Dirac PM states forming pairs of pseudo-Dirac fields. ($g$) Loops of PM and $W_I$ gauge bosons can realize potentially important dark dipole moment-like 
couplings of the SM fermions to the DP, making possibly substantial alterations in the associated phenomenology, as suggested in previous work. 
($h$) The non-hermitian, $W_R$ and $W_I$ gauge bosons have properties which are semi-quantitatively not too dissimilar from those encountered in the usual LRM and in the simpler scenario 
explored in Ref.\cite{Rueter:2019wdf} where the important mixing of the DP with the hermitian combination $W_I+W_I^\dagger$ was previously noted.  However, due to the mixing of the SM 
and PM fields at the $\sim 10^{-(3-4)}$ level some novel and yet to be explored new effects are possible. ($i$) The two new heavy neutral gauge bosons present in this setup, 
$Z_{R,I}$, generally 
undergo substantial mixing into the $Z_{1,2}$ mass eigenstates, one of which is always heavier (lighter) than the corresponding pure LRM $Z_R$ `reference' state with generally 
stronger (weaker) couplings given the same input parameter values. Making some reasonable model assumptions for purposes of demonstration, estimates were obtained for the lower 
bounds on the masses of both of these states from existing ATLAS dilepton resonance search data and then these reaches were extrapolated to obtain the corresponding mass reaches for the 
100 TeV FCC-hh under an identical set of assumptions. 

The extension of the SM gauge group to the LRM in addition to the existence of a non-abelian symmetry for the dark sector provides a phenomenologically rich and interesting direction to 
explore in our search for a more UV-complete model of the gauge interactions of the visible and dark sectors. Further steps in this direction will be taken in future work.

\section*{Acknowledgements}
The author would like to particularly thank J.L. Hewett and G. Wojcik for valuable discussions during the early aspects of this work.  This work was supported by the 
Department of Energy, Contract DE-AC02-76SF00515.




\begin{thebibliography}{99}

\bibitem{Planck:2018vyg}
N.~Aghanim \textit{et al.} [Planck],
Astron. Astrophys. \textbf{641}, A6 (2020)
[erratum: Astron. Astrophys. \textbf{652}, C4 (2021)]
[arXiv:1807.06209 [astro-ph.CO]].

\bibitem{Kawasaki:2013ae} 
  M.~Kawasaki and K.~Nakayama,
  Ann.\ Rev.\ Nucl.\ Part.\ Sci.\  {\bf 63}, 69 (2013)
  [arXiv:1301.1123 [hep-ph]].
 
\bibitem{Graham:2015ouw} 
  P.~W.~Graham, I.~G.~Irastorza, S.~K.~Lamoreaux, A.~Lindner and K.~A.~van Bibber,
  Ann.\ Rev.\ Nucl.\ Part.\ Sci.\  {\bf 65}, 485 (2015)
  [arXiv:1602.00039 [hep-ex]].

\bibitem{Irastorza:2018dyq}
I.~G.~Irastorza and J.~Redondo,
Prog. Part. Nucl. Phys. \textbf{102}, 89-159 (2018)
[arXiv:1801.08127 [hep-ph]].

\bibitem{Arcadi:2017kky} 
 G.~Arcadi, M.~Dutra, P.~Ghosh, M.~Lindner, Y.~Mambrini, M.~Pierre, S.~Profumo and F.~S.~Queiroz, 
Eur. Phys. J. C \textbf{78}, no.3, 203 (2018)
[arXiv:1703.07364 [hep-ph]].
  
\bibitem{Roszkowski:2017nbc}
L.~Roszkowski, E.~M.~Sessolo and S.~Trojanowski,
Rept. Prog. Phys. \textbf{81}, no.6, 066201 (2018)
[arXiv:1707.06277 [hep-ph]].
 
\bibitem{LHC}
  K. Pachal,  ``Dark Matter Searches at ATLAS and CMS'', given at the $8^{th}$ {\it {Edition of the Large Hadron Collider Physics Conference}}, 25-30 May, 2020.
 
\bibitem{Aprile:2018dbl}
E.~Aprile \textit{et al.} [XENON],
Phys. Rev. Lett. \textbf{121}, no.11, 111302 (2018)
[arXiv:1805.12562 [astro-ph.CO]]. 

\bibitem{Fermi-LAT:2016uux}
A.~Albert \textit{et al.} [Fermi-LAT and DES],
Astrophys. J. \textbf{834}, no.2, 110 (2017)
[arXiv:1611.03184 [astro-ph.HE]].

\bibitem{Amole:2019fdf}
C.~Amole \textit{et al.} [PICO],
Phys. Rev. D \textbf{100}, no.2, 022001 (2019)
[arXiv:1902.04031 [astro-ph.CO]].

\bibitem{LZ:2022ufs}
J.~Aalbers \textit{et al.} [LZ],
[arXiv:2207.03764 [hep-ex]].


\bibitem{Alexander:2016aln} 
  J.~Alexander {\it et al.},
  arXiv:1608.08632 [hep-ph].

\bibitem{Battaglieri:2017aum} 
  M.~Battaglieri {\it et al.},
  arXiv:1707.04591 [hep-ph].
  
\bibitem{Bertone:2018krk}
G.~Bertone and T.~Tait, M.P.,
Nature \textbf{562}, no.7725, 51-56 (2018)
[arXiv:1810.01668 [astro-ph.CO]].

\bibitem{Cooley:2022ufh}
J.~Cooley, T.~Lin, W.~H.~Lippincott, T.~R.~Slatyer, T.~T.~Yu, D.~S.~Akerib, T.~Aramaki, D.~Baxter, T.~Bringmann and R.~Bunker, \textit{et al.}
[arXiv:2209.07426 [hep-ph]].

\bibitem{Boveia:2022syt}
A.~Boveia, T.~Y.~Chen, C.~Doglioni, A.~Drlica-Wagner, S.~Gori, W.~H.~Lippincott, M.~E.~Monzani, C.~Prescod-Weinstein, B.~Shakya and T.~R.~Slatyer, \textit{et al.}
[arXiv:2210.01770 [hep-ph]].


\bibitem{Schuster:2021mlr}
P.~Schuster, N.~Toro and K.~Zhou,
Phys. Rev. D \textbf{105}, no.3, 035036 (2022)
doi:10.1103/PhysRevD.105.035036
[arXiv:2112.02104 [hep-ph]].


\bibitem{KM}
  B.~Holdom,
  Phys.\ Lett.\  {\bf 166B}, 196 (1986) and
  Phys.\ Lett.\ B {\bf 178}, 65 (1986); 
  K.~R.~Dienes, C.~F.~Kolda and J.~March-Russell,
  Nucl.\ Phys.\ B {\bf 492}, 104 (1997)
  [hep-ph/9610479];
  F.~Del Aguila,
  Acta Phys.\ Polon.\ B {\bf 25}, 1317 (1994)
  [hep-ph/9404323];
  K.~S.~Babu, C.~F.~Kolda and J.~March-Russell,
  Phys.\ Rev.\ D {\bf 54}, 4635 (1996)
  [hep-ph/9603212];
  T.~G.~Rizzo,
  Phys.\ Rev.\ D {\bf 59}, 015020 (1998)
  [hep-ph/9806397].

\bibitem{vectorportal} 
 There has been a huge amount of work on this subject; see, for example, 
  D.~Feldman, B.~Kors and P.~Nath,
  Phys.\ Rev.\ D {\bf 75}, 023503 (2007)
  [hep-ph/0610133];
  D.~Feldman, Z.~Liu and P.~Nath,
  Phys.\ Rev.\ D {\bf 75}, 115001 (2007)
  [hep-ph/0702123 [HEP-PH]].;
  M.~Pospelov, A.~Ritz and M.~B.~Voloshin,
  Phys.\ Lett.\ B {\bf 662}, 53 (2008)
  [arXiv:0711.4866 [hep-ph]];
  M.~Pospelov,
  Phys.\ Rev.\ D {\bf 80}, 095002 (2009)
  [arXiv:0811.1030 [hep-ph]]; 
  H.~Davoudiasl, H.~S.~Lee and W.~J.~Marciano,
  Phys.\ Rev.\ Lett.\  {\bf 109}, 031802 (2012)
  [arXiv:1205.2709 [hep-ph]] and 
  Phys.\ Rev.\ D {\bf 85}, 115019 (2012)
  doi:10.1103/PhysRevD.85.115019
  [arXiv:1203.2947 [hep-ph]];
  R.~Essig {\it et al.},
  arXiv:1311.0029 [hep-ph];
  E.~Izaguirre, G.~Krnjaic, P.~Schuster and N.~Toro,
  Phys.\ Rev.\ Lett.\  {\bf 115}, no. 25, 251301 (2015)
  [arXiv:1505.00011 [hep-ph]];
  M.~Khlopov,
  Int.\ J.\ Mod.\ Phys.\ A {\bf 28}, 1330042 (2013)
  [arXiv:1311.2468 [astro-ph.CO]];
 For a general overview and introduction to this framework, see  
  D.~Curtin, R.~Essig, S.~Gori and J.~Shelton,
  JHEP {\bf 1502}, 157 (2015)
  [arXiv:1412.0018 [hep-ph]].
  

\bibitem{Gherghetta:2019coi}
T.~Gherghetta, J.~Kersten, K.~Olive and M.~Pospelov,
Phys. Rev. D \textbf{100}, no.9, 095001 (2019)
[arXiv:1909.00696 [hep-ph]].


\bibitem{Steigman:2015hda} 
  G.~Steigman,
  Phys.\ Rev.\ D {\bf 91}, no. 8, 083538 (2015)
  [arXiv:1502.01884 [astro-ph.CO]].
  
\bibitem{Saikawa:2020swg}
K.~Saikawa and S.~Shirai,
[arXiv:2005.03544 [hep-ph]].
 
\bibitem{Fabbrichesi:2020wbt}
M.~Fabbrichesi, E.~Gabrielli and G.~Lanfranchi,
[arXiv:2005.01515 [hep-ph]].


\bibitem{Graham:2021ggy}
M.~Graham, C.~Hearty and M.~Williams,
[arXiv:2104.10280 [hep-ph]].


\bibitem{Rizzo:2018vlb}
T.~G.~Rizzo,
Phys. Rev. D \textbf{99}, no.11, 115024 (2019)
[arXiv:1810.07531 [hep-ph]].

\bibitem{Rueter:2019wdf}
T.~D.~Rueter and T.~G.~Rizzo,
Phys. Rev. D \textbf{101}, no.1, 015014 (2020)
[arXiv:1909.09160 [hep-ph]].

\bibitem{Kim:2019oyh}
J.~H.~Kim, S.~D.~Lane, H.~S.~Lee, I.~M.~Lewis and M.~Sullivan,
Phys. Rev. D \textbf{101}, no.3, 035041 (2020)
[arXiv:1904.05893 [hep-ph]].

\bibitem{Rueter:2020qhf}
T.~D.~Rueter and T.~G.~Rizzo,
[arXiv:2011.03529 [hep-ph]].

\bibitem{Wojcik:2020wgm}
G.~N.~Wojcik and T.~G.~Rizzo,
Phys. Rev. D \textbf{105}, no.1, 015032 (2022)
[arXiv:2012.05406 [hep-ph]].

\bibitem{Rizzo:2021lob}
T.~G.~Rizzo,
JHEP \textbf{11}, 035 (2021)
[arXiv:2106.11150 [hep-ph]].

\bibitem{Rizzo:2022qan}
T.~G.~Rizzo,
[arXiv:2202.02222 [hep-ph]].

\bibitem{Wojcik:2022rtk}
G.~N.~Wojcik,
[arXiv:2205.11545 [hep-ph]].

\bibitem{Rizzo:2022jti}
T.~G.~Rizzo,
[arXiv:2206.09814 [hep-ph]].

\bibitem{Rizzo:2022lpm}
T.~G.~Rizzo,
Phys. Rev. D \textbf{106}, no.9, 095024 (2022)
[arXiv:2209.00688 [hep-ph]].

\bibitem{Wojcik:2022woa}
G.~N.~Wojcik, L.~L.~Everett, S.~T.~Eu and R.~Ximenes,
[arXiv:2211.09918 [hep-ph]].

\bibitem{Carvunis:2022yur}
A.~Carvunis, N.~McGinnis and D.~E.~Morrissey,
[arXiv:2209.14305 [hep-ph]].

\bibitem{Verma:2022nyd}
S.~Verma, S.~Biswas, A.~Chatterjee and J.~Ganguly,
[arXiv:2209.13888 [hep-ph]].


\bibitem{Slatyer:2015jla}
T.~R.~Slatyer,
Phys. Rev. D \textbf{93}, no.2, 023527 (2016)
[arXiv:1506.03811 [hep-ph]].

\bibitem{Liu:2016cnk} 
  H.~Liu, T.~R.~Slatyer and J.~Zavala,
  Phys.\ Rev.\ D {\bf 94}, no. 6, 063507 (2016)
  [arXiv:1604.02457 [astro-ph.CO]].

\bibitem{Leane:2018kjk}
R.~K.~Leane, T.~R.~Slatyer, J.~F.~Beacom and K.~C.~Ng,
Phys. \ Rev. \ D \textbf{98}, no.2, 023016 (2018)
[arXiv:1805.10305 [hep-ph]].

\bibitem{Bauer:2022nwt}
For related work on the possibilities of KM and DM physics employing this same gauge group, see 
M.~Bauer and P.~Foldenauer,
Phys. Rev. Lett. \textbf{129}, no.17, 171801 (2022)
[arXiv:2207.00023 [hep-ph]].

\bibitem{e6}
See, for example, 
F.~Gursey, P.~Ramond and P.~Sikivie,
Phys. Lett. B \textbf{60}, 177-180 (1976); 
Y.~Achiman and B.~Stech,
Phys. Lett. B \textbf{77}, 389-393 (1978); 
Q.~Shafi,
Phys. Lett. B \textbf{79}, 301-303 (1978).

\bibitem{Hewett:1988xc}
J.~L.~Hewett and T.~G.~Rizzo,
Phys. Rept. \textbf{183}, 193 (1989).

\bibitem{Pati:1974yy}
J.~C.~Pati and A.~Salam,
Phys. Rev. D \textbf{10}, 275-289 (1974)
[erratum: Phys. Rev. D \textbf{11}, 703-703 (1975)].

\bibitem{Mohapatra:1974hk}
R.~N.~Mohapatra and J.~C.~Pati,
Phys. Rev. D \textbf{11}, 566-571 (1975)

\bibitem{Mohapatra:1974gc}
R.~N.~Mohapatra and J.~C.~Pati,
Phys. Rev. D \textbf{11}, 2558 (1975)

\bibitem{Senjanovic:1975rk}
G.~Senjanovic and R.~N.~Mohapatra,
Phys. Rev. D \textbf{12}, 1502 (1975)

\bibitem{book}
R. N.Mohapatra, (1986), 10.1007/978-1-4757-1928-4.

\bibitem{Dutka:2022lug}
For a recent analysis of this mass scale, see for example, T.~P.~Dutka and J.~Gargalionis,
[arXiv:2211.02054 [hep-ph]].

\bibitem{Rizzo:2006nw}
See, for example, 
T.~G.~Rizzo,
[arXiv:hep-ph/0610104 [hep-ph]].


\bibitem{GBET}
  M.~S.~Chanowitz and M.~K.~Gaillard,
  Nucl.\ Phys.\ B {\bf 261}, 379 (1985);
  B.~W.~Lee, C.~Quigg and H.~B.~Thacker,
  Phys.\ Rev.\ D {\bf 16}, 1519 (1977);
  J.~M.~Cornwall, D.~N.~Levin and G.~Tiktopoulos,
  Phys.\ Rev.\ D {\bf 10}, 1145 (1974)
  Erratum: [Phys.\ Rev.\ D {\bf 11}, 972 (1975)];
  G.~J.~Gounaris, R.~Kogerler and H.~Neufeld,
  Phys.\ Rev.\ D {\bf 34}, 3257 (1986).
 

\bibitem{quasi}
Such heavy neutral lepton states have been discussed in a number of different contexts; see for example, 
A.~Das, P.~S.~Bhupal Dev and N.~Okada,
Phys. Lett. B \textbf{735}, 364-370 (2014)
[arXiv:1405.0177 [hep-ph]]; 
A.~de Gouvea, W.~C.~Huang and J.~Jenkins,
Phys. Rev. D \textbf{80}, 073007 (2009)
[arXiv:0906.1611 [hep-ph]];
G.~Anamiati, M.~Hirsch and E.~Nardi,
JHEP \textbf{10}, 010 (2016)
[arXiv:1607.05641 [hep-ph]];
P.~Hern\'andez, J.~Jones-P\'erez and O.~Suarez-Navarro,
Eur. Phys. J. C \textbf{79}, no.3, 220 (2019)
[arXiv:1810.07210 [hep-ph]];
D.~Chang and O.~C.~W.~Kong,
Phys. Lett. B \textbf{477}, 416-423 (2000)
[arXiv:hep-ph/9912268 [hep-ph]];
S.~Bahrami, M.~Frank, D.~K.~Ghosh, N.~Ghosh and I.~Saha,
Phys. Rev. D \textbf{95}, no.9, 095024 (2017)
[arXiv:1612.06334 [hep-ph]].



\bibitem{ATLAS:2019erb}
G.~Aad \textit{et al.} [ATLAS],
Phys. Lett. B \textbf{796}, 68-87 (2019)
[arXiv:1903.06248 [hep-ex]].

\bibitem{Helsens:2019bfw}
C.~Helsens, D.~Jamin, M.~L.~Mangano, T.~G.~Rizzo and M.~Selvaggi,
Eur. Phys. J. C \textbf{79}, 569 (2019)
[arXiv:1902.11217 [hep-ph]].


\bibitem{wpsearch}
See, for example, 
A.~M.~Sirunyan \textit{et al.} [CMS],
Phys. Lett. B \textbf{820}, 136535 (2021)
[arXiv:2104.04831 [hep-ex]];
G.~Aad \textit{et al.} [ATLAS],
JHEP \textbf{03}, 145 (2020)
[arXiv:1910.08447 [hep-ex]];
A.~Tumasyan \textit{et al.} [CMS],
JHEP \textbf{04}, 047 (2022)
[arXiv:2112.03949 [hep-ex]];
G.~Aad \textit{et al.} [ATLAS],
Phys. Rev. D \textbf{100}, no.5, 052013 (2019)
[arXiv:1906.05609 [hep-ex]];
A.~Tumasyan \textit{et al.} [CMS],
JHEP \textbf{07}, 067 (2022)
[arXiv:2202.06075 [hep-ex]];
A.~M.~Sirunyan \textit{et al.} [CMS],
JHEP \textbf{05}, 033 (2020)
[arXiv:1911.03947 [hep-ex]];
 ATLAS Collaboration,
ATLAS-CONF-2021-043.

\bibitem{ATLAS:2022jsi}
 ATLAS Collaboration, 
``Combination of searches for heavy resonances using 139 fb$^{-1}$ of proton\textendash{}proton collision data at $\sqrt{s}$ = 13 TeV with the ATLAS detector,''
ATLAS-CONF-2022-028.

\bibitem{CDF:2022hxs}
T.~Aaltonen \textit{et al.} [CDF],
``High-precision measurement of the $W$  boson mass with the CDF II detector,''
Science \textbf{376}, no.6589, 170-176 (2022).




\end{thebibliography}
\end{document}